\documentclass[a4paper, amsfonts, amssymb, amsmath, reprint, showkeys, nofootinbib, twoside, aps]{revtex4-1}
\usepackage[english]{babel}
\usepackage[utf8]{inputenc}

\usepackage[colorinlistoftodos, color=green!40, prependcaption]{todonotes}
\usepackage{amsthm}
\usepackage{mathtools}
\usepackage{physics}
\usepackage{xcolor}
\usepackage{graphicx}
\usepackage[left=23mm,right=13mm,top=35mm,columnsep=15pt]{geometry} 
\usepackage{adjustbox}
\usepackage{placeins}
\usepackage[T1]{fontenc}
\usepackage{lipsum}
\usepackage{csquotes}

\usepackage[pdftex, pdftitle={Article}, pdfauthor={Author}]{hyperref}
\begin{document}
\renewcommand{\figurename}{FIG.}

\title{Machine Learning Green’s Functions of Strongly Correlated Hubbard Models}

\author{Mateo Cárdenes Wuttig}
    \email[]{mateo.cardeneswuttig@yale.edu}
    \affiliation{Department of Chemistry, Yale University, New
Haven, Connecticut 06520, United States}
    \affiliation{Department of Applied Physics, Yale University, New
Haven, Connecticut 06520, United States}

\date{\today}

\begin{abstract}
We demonstrate that a machine learning framework based on kernel ridge regression can encode and predict the self-energy of one-dimensional Hubbard models using only mean-field features such as static and dynamic Hartree-Fock quantities and first-order \textit{GW} calculations. This approach is applicable across a wide range of on-site Coulomb interaction strengths $U/t$, ranging from weakly interacting systems ($U/t \ll 1$) to strong correlations ($U/t > 8$). The predicted self-energy is transformed via Dyson’s equation and analytic continuation to obtain the real-frequency Green's function, which allows access to the spectral function and density of states. This method can be used for nearest-neighbor interactions $t$ and long-range hopping terms $t'$, $t''$, and $t'''$. 
\end{abstract}

\maketitle

\section{Introduction} \label{sec:outline}
In this article, we extend a previously introduced machine learning (ML) framework developed for molecules\cite{doi:10.1021/acs.jctc.3c01146} to strongly correlated electronic systems, characterized by suppressed quasi-particle peaks, large self-energy, and broad bands\cite{RevModPhys.68.13, 10.1063/1.1712502, RevModPhys.78.865, PhysRevB.102.081110}. For weakly interacting systems, one can obtain an accurate mean-field solution through Hartree-Fock (HF) calculations, where the quantum many-body wavefunction is approximated as a single Slater determinant\cite{PhysRev.81.385,doi.org/10.1007/BF02188656}, neglecting quantum fluctuations. Density-functional theory (DFT) enables accurate and efficient computations based on the electron density\cite{PhysRev.140.A1133}, although the exact energy functional remains unknown, and approximate exchange–correlation functionals lead to systematic errors\cite{doi:10.1021/cr200107z}. Spectral properties of systems where electron correlations must be considered explicitly, however, require numerically intensive many-body methods that are typically limited to small or low-dimensional systems\cite{PhysRevB.88.041107,PhysRevB.91.155107,PhysRevB.100.115154}, such as many-body perturbation theory (\textit{GW})\cite{10.3389/fchem.2019.00377} or coupled-cluster (CC) theory\cite{https://doi.org/10.1002/qua.560440808,RevModPhys.79.291}.

Here, we demonstrate an ML model that learns the imaginary-frequency self-energy function from mean-field input features and generalizes to previously unseen model parameters, providing many-body spectral predictions based on mean-field data. Our work serves as a proof-of-concept for ML methods that bridge the gap between mean-field calculations at the Hartree-Fock level and approximate many-body methods. We study the one-dimensional half-filled Hubbard model with varying on-site repulsion $U$. This system is well-understood\cite{Essler_Frahm_Göhmann_Klümper_Korepin_2005, PhysRevLett.73.732, PhysRevB.109.045102}, making it an ideal candidate to investigate the quality of ML predictions. We use the many-body Green’s function formalism\cite{PhysRev.139.A796} to calculate spectral properties\cite{PhysRevB.90.155136}. A large variety of methods, such as \textit{GW} or CC, can be expressed in this framework, which may allow our ML architecture to be integrated into existing workflows.

We note that a large number of ML approaches have been used to simulate correlated quantum systems. Many methods, such as restricted Boltzmann machines\cite{PhysRevB.100.245123, PhysRevB.96.205152}, deep neural networks\cite{PhysRevB.102.205122}, and Transformer architectures\cite{10.1145/3581784.3607061}, commonly described as Neural Quantum States (NQS)\cite{Lange_2024}, have been particularly successful in describing ground-state properties. The accurate prediction of excited-state observables and spectral quantities using ML methods, however, remains more challenging, as these require resolving frequency-dependent many-body correlations and dynamical information beyond ground-state wavefunctions\cite{PhysRevB.90.155136,PhysRevB.100.245123, PhysRevLett.131.046501, PhysRevB.105.205130, PhysRevB.103.245118, Zhu_2025}. Recent deep-learning approaches have demonstrated promising performance for quasiparticle excitations within the GW formalism\cite{https://doi.org/10.1038/s41467-024-53748-7}, although such methods often rely on large training datasets or achieve high accuracy only within specific parameter regimes or optimization schemes\cite{PhysRevResearch.6.043280}.

On the other hand, there exists a large variety of methods that can find highly accurate descriptions of many-body ground states, including tensor networks\cite{PhysRevLett.69.2863, SCHOLLWOCK201196}, Quantum Monte Carlo\cite{PhysRevB.80.075116}, and NQS\cite{doi:10.1126/science.aag2302,doi.org/10.1038/s41467-020-15724-9, PhysRevLett.121.167204}.
These methods, however, do not allow for a precise computation of spectral quantities. Additionally, some methods, such as tensor networks, are particularly well-suited for (quasi) one-dimensional systems\cite{annurev:/content/journals/10.1146/annurev-conmatphys-020911-125018} and struggle with long-range interactions\cite{RevModPhys.77.259} or periodic boundary conditions\cite{PhysRevLett.93.227205}, while others, like Quantum Monte Carlo, suffer from the sign problem with increasing Coulomb repulsion\cite{PhysRevB.41.9301}.

Our approach consists of using a Kernel Ridge Regression (KRR) architecture\cite{https://doi.org/10.1002/qua.24939} to learn and predict the self-energy on a compact imaginary frequency grid. We can combine this quantity with the mean-field Green's function via Dyson's equation to obtain the many-body Green's function. The real-frequency spectral quantities, such as the spectral function and density of states (DOS), can be obtained via analytic continuation (A.C.) from the functions calculated on the imaginary axis\cite{fetter2003quantum}. 

This work provides a route to ML-based methods that may offer several advantages in comparison to existing numerical methods. 
First, our method is accurate for short-range nearest-neighbor interactions as well as long-range hoppings. Additionally, there is no structural difference between open and periodic boundary conditions, and the framework maintains consistent accuracy even for long-range interactions. We investigate different hopping parameters $t, t', t''$ and $t'''$ between up to four neighbors, all of which are realized in different condensed matter systems, making our method a versatile approach for the study of spectral properties in real materials. Lastly, our method works well over a broad range of interaction strengths $U/t$ from weak to strong correlations. In addition, the KRR model is fully interpretable, increasing confidence in the results, which makes it stand out compared to most other ML architectures. This is of major advantage in understanding how mean-field quantities are related to many-body correlations. 

The article is structured as follows. In section \ref{sec:methods}, we introduce the ML architecture and workflow and explain our metric of accuracy. In section \ref{sec:nn}, we demonstrate that our ML framework can encode the self-energy of a nearest-neighbor Hubbard model for a wide range of Coulomb repulsions $U$, reproducing the training data and generalizing to unseen test data. We argue that the differences between the exact density of states and the ML prediction mainly stem from the analytic continuation. To prove this, we present an extensive overview of the absolute relative difference between the predicted and exact self-energy. Then, we show in section \ref{sec:longrange} that our ML model also allows for accurate predictions of systems with long-range hopping terms $t'$, $t''$, and $t'''$. In section \ref{sec:conclusion}, we discuss our results.
 
\section{Methods}\label{sec:methods}
\subsection{1d Hubbard Model}
\begin{figure}[htbp!]
 \centering
 \begin{adjustbox}{center}
   \includegraphics[width=0.66\columnwidth]{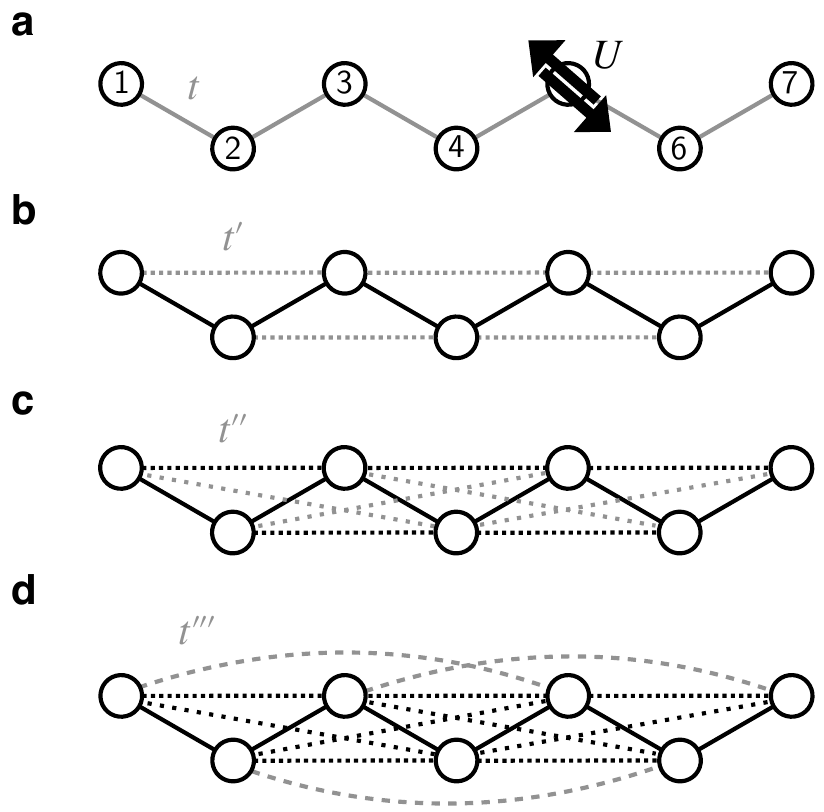}
 \end{adjustbox}
 \caption{One-dimensional Hubbard model with on-site Coulomb repulsion $U$. (a) Nearest-neighbor hopping $t$. (b) Zig-zag ladder with next-neighbor hopping $t'$. (c) Zig-zag ladder with diagonal hopping term $t''$. (d) Additional hopping $t'''$ between next-nearest-neighbor atoms on both legs of the zig-zag ladder.}
 \label{fig:figS1}
\end{figure}

We consider a one-dimensional Hubbard model of length $L=10$ with one orbital per site and periodic boundary conditions at half-filling, described by the following Hamiltonian
\begin{equation}
    H = - \sum_{\langle ij \rangle, \sigma} t_{ij} (c^\dag_{i\sigma}c_{j\sigma} + \text{h.c.}) + U \sum_{i} n_{i\uparrow} n_{i\downarrow}
\end{equation}
where $c^\dag_{i\sigma}$ ($c_{i\sigma}$) creates (destroys) an electron of spin $\sigma = \,\uparrow, \downarrow$ on site $i = 1,...,L$, and $n_{i\sigma} = c^{\dag}_{i\sigma}c_{i\sigma}$ counts the number of electrons. We sum over all bonds $\langle ij \rangle$ and consider nearest-neighbor interactions $t_{ij} = t$ for $\abs{i-j}=1$ and long-range interactions $t_{ij} =t^{(n)} \delta_{\abs{i-j},n}$ between two ($t' = t^{(2)}$), three ($t'' = t^{(3)}$), and four ($t''' = t^{(4)}$) neighbors. In fig. \ref{fig:figS1}, we present these systems, including a Hubbard chain with nearest neighbor interactions $t$ (a); next-nearest neighbor hopping $t'$ (b), which resembles a zig-zag ladder; next-next-nearest neighbor hopping $t''$ (c), equivalent to a zig-zag ladder with diagonal interactions; and also a $t'''$-hopping (d), corresponding to a folded zig-zag ladder with an armchair structure, where $t'''$ describes hopping between two layers of zig-zag ladders. All parameters are given in units of $t=1$. 

\subsection{Many-Body Green's Function}
The single-particle Green’s function describes how an electron propagates from site $j$ at time $t_0 = 0$ to site $i$ at time $t > 0$
\begin{equation}
    G(t)^{\sigma}_{ij} = -\mathrm{i} \theta(t) \expval{c_{i\sigma}(t) c_{j\sigma}^{\dagger} (t_0)}{\psi_0} \,.
\end{equation}
Here, $\ket{\psi_0}$ denotes the ground state wave function at zero temperature. We only consider the retarded Green's function, and we fix $\sigma = \,\uparrow$ and omit the spin index.
In the frequency domain, the addition part of the Green's function is defined as
\begin{equation}\label{eq:greensfunction_frequency}
    G(z)^{\text{EA}}_{ij} = \expval{c_{i\sigma} \frac{1}{z - (\hat{H} - E)} c_{j\sigma}^{\dagger}}{\psi_0}
\end{equation}
such that $G(z)= G(z)^{\text{EA}} + G(z)^{\text{IP}}$ is the sum of the removal (IP) and addition (EA) parts.
$E$ is the ground state energy determined by the Hamiltonian $\hat{H}$. 
We can express the propagator both on the real frequency axis $z = \omega + \mathrm{i} \eta$ as a function of energy with a broadening factor $\eta$, or on the imaginary frequency axis $z = E_F + \mathrm{i} \omega$, where $E_F$ is the Fermi energy. We know that $E_F = U/2$ at half-filling for the nearest-neighbor Hubbard model, and we present the density of states (DOS) as a function of $E - U/2$. 
We use a discrete set of $N_{\omega_k} = 30$ imaginary frequencies $\{\mathrm{i}\omega_k\}$ with $k=1,...,N_{\omega_k}$ based on a modified Gauss-Legendre grid\cite{doi:10.1021/acs.jctc.3c01146} to describe the Green’s function. Hence, $G(\omega)$ is a set of $N_{\omega_k}$ matrices of shape $N_{AO} \times N_{AO}$, where the number of atomic orbitals is $N_{AO} = L$. 

We can relate the non-interacting mean-field Green's function $G_0$, obtained from Hartree-Fock (HF) calculations, and the self-energy $\Sigma$ to the many-body Green's function $G$ via Dyson's equation\cite{fetter2003quantum}:
\begin{equation}\label{eq:dysoneq}
    G^{-1}(\omega) = G^{-1}_0(\omega) - \Sigma(\omega)  \,.
\end{equation}
Thus, we only have to know one of ($G$, $\Sigma$) to obtain the many-body spectral properties for a given $G_0$. Note that all three objects have the same shape $N_{\omega_k} \times N_{AO} \times N_{AO}$, where eq. \ref{eq:dysoneq} is a matrix equation.
The many-body Green's function $G$ on the real axis gives access to the spectral function
\begin{equation}
    A(\omega) = -\frac{\Im[G(\omega)]}{\pi}
\end{equation}
and the density of states
\begin{equation}
    \text{DOS}(\omega) = \Tr[A(\omega)] \,.
\end{equation}

In our workflow, we predict the self-energy on a discrete imaginary frequency axis $\Sigma(\{\text{i}\omega_k\})$ since this function is smooth and thus an easier target for the ML model\cite{doi:10.1021/acs.jctc.3c01146}. We use Padé analytic continuation\cite{doi.org/10.1007/BF00655090} to transform the self-energy from the imaginary frequency axis to the real energy axis $\Sigma(\{\text{i}\omega_k\}) \rightarrow \Sigma(\{E_k\} - U/2)$ with a broadening factor of $\eta = 0.015
$. We use a linear grid of 400 energy points \hbox{$\{E_k\} \in [-2, 2]$ eV} for $U/t < 4$, \hbox{$\{E_k\} \in [-6, 6]$ eV} for $U/t < 8$, and \hbox{$\{E_k\} \in [-10, 10]$ eV} otherwise. All HF calculations are performed with an iterative algorithm that gradually introduces random noise to the initial guess of the density matrix until the calculation converges. The converged Restricted Hartree-Fock (RHF) state is used to calculate $G_0$. Note that for $U=0$, the Green's function $G_0$ can easily be obtained from exact diagonalization. The exact Green's function $G$ is computed via Full-Configuration Interaction (FCI). We are working in a basis of fixed particle number, leading to a total Hilbert space dimension of $\dim{H} = \binom{L}{N_{\uparrow}} \binom{L}{N_{\downarrow}}$, where $N_{\uparrow} = N_{\downarrow} = L/2$. All calculations are performed at zero temperature. Both $G$ and $G_0$ are computed in molecular orbital (M.O.) basis and are transformed to atomic orbital (A.O.) basis. We perform all ML predictions in A.O. basis to benefit from the translation invariance symmetry of the Hamiltonian. 
We use the PySCF package\cite{10.1063/5.0006074, https://doi.org/10.1002/wcms.1340} and the FCDMFT package\cite{PhysRevX.11.021006, 10.1021/acs.jctc.9b00934, PhysRevB.100.115154} for Green's function and \textit{GW} calculations\cite{doi:10.1021/acs.jctc.0c00704}.

\subsection{Machine-Learning Workflow}
In the following, we refer to the exact many-body Green's function computed by FCI as FCI-GF, to the mean-field Green's function computed by RHF as HF-GF, and to the predicted Green's function as ML-GF. 
Our workflow is based on a previous work by Venturella et al.\cite{doi:10.1021/acs.jctc.3c01146} For each Hamiltonian defined by the hopping parameters $(t,t',t'',t''')$, we define separate KRR models $f$ for on- and off-diagonal elements as well as the real and imaginary parts of every frequency value $\omega_k$. These $4 N_{\omega_k}$ independent models are trained with data for different values of $U$. For all calculations, we provide 604 FCI-GF as training data and 21 FCI-GF as test data with mutually exclusive Coulomb repulsion values between $U = 0.25$ and $U=10$. Both test and training data are calculated for the same Hamiltonian parameters $(t,t',t'',t''')$. Our ML model $f^{\text{on/off},\text{Re/Im}, \omega_k}$ learns by using the self-energy $\Sigma(\omega_k)^{\text{training}}_{ij}$, obtained from the FCI-GF training data via eq. \ref{eq:dysoneq}, as a target with features $\mathbf{f}^{\text{training}}_{ij}$
\begin{equation}\label{eq:learning}
    f^{\text{on/off},\text{Re/Im}, \omega_k}( \mathbf{f}^{\text{training}}_{ij}) \rightarrow \text{Re/Im}[\Sigma(\omega_k)^{\text{training}}_{ij}] \,.
\end{equation} 
Once trained, the model can predict a self-energy based on features of unseen test data $\mathbf{f}^{\text{test}}_{ij}$
\begin{equation}\label{eq:prediction}
    \text{Re/Im}[\Sigma(\omega_k)^{\text{pred}}_{ij}] = f^{\text{on/off},\text{Re/Im}, \omega_k}(\mathbf{f}^{\text{test}}_{ij})
\end{equation}
which can be converted to the ML-GF $G^{\text{pred}}$ via Dyson's equation by combining \hbox{$\Sigma^\text{pred} = \Re(\Sigma^\text{pred}) + \mathrm{i} \Im(\Sigma^\text{pred})$} with $G_0$. The feature vector $\mathbf{f}_{ij}$ is independent of frequency $\omega_k$ and contains the following 54 elements
\begin{equation}\label{eq:feature}
\begin{split}
        \mathbf{f}_{ij} = & \Big(U, h_{ij}, F_{ij}, J_{ij}, K_{ij}, \rho_{ij}, \\ & \,\{\Re[G_{0}(\omega_n)_{ij}]\}_{n=1,...,2\cdot N_{\omega_n}}, \{\Im[G_{0}(\omega_n)_{ij}]\}, \\
        & \,\{\Re[\Delta(\omega_n)_{ij}]\}_{n=1,...,N_{\omega_n}}, \{\Im[\Delta(\omega_n)_{ij}]\}, \\ &
        \,\{\Re[\Sigma_{\text{GW}}(\omega_n)_{ij}]\}, \{\Im[\Sigma_{\text{GW}}(\omega_n)_{ij}]\}\Big) \,.
\end{split}
\end{equation}
We include the on-site repulsion $U$ and the matrix elements $(i,j)$ of the core Hamiltonian matrix $h$, the Fock matrix $F$, the Coulomb matrix $J$, the exchange matrix $K$, and the one-electron reduced density matrix $\rho$. Although the Fock matrix $F$ is linearly related to the static features $(h,J,K)$, including it serves as a form of feature augmentation and may improve ML performance by stabilizing the regression.
Additionally, we provide the respective real and imaginary matrix elements of the HF-GF $G_{0}(\omega_n)$ and the hybridization function $\Delta(\omega_n)$\cite{doi:10.1021/acs.jctc.3c01146, doi.org/10.1038/s43588-025-00810-z, PhysRevX.11.021006, 10.1021/acs.jctc.9b00934}, defined separately for on- and off-diagonal elements
\begin{equation}
\begin{split}
    \Delta(\omega)_{ij} = & E_F  + \mathrm{i} \omega - F_{ij} + \\ &
\begin{cases}
    (G_0(\omega)^{-1})_{ii}\,, & i=j \\
    \frac{G_0(\omega)_{ij}}{G_0(\omega)_{ij}G_0(\omega)_{ji} - G_0(\omega)_{ii}G_0(\omega)_{jj}}\,, & i\neq j
\end{cases} 
\end{split}
\end{equation}
where $E_F$ is determined from HF calculations. We also include a first order $G_0 W_0$ calculation for the self-energy $\Sigma_{\text{GW}}(\omega_n)$. Note that $\Delta(\omega_n)$ and $\Sigma_{\text{GW}}(\omega_n)$ are computed for $N_{\omega_n} = 6$ imaginary frequency values from a modified Gauss-Legendre grid, while $G_{0}(\omega_n)$ is computed on a denser grid of $2\cdot N_{\omega_n} = 12$ imaginary frequencies. All features are transformed from M.O. to A.O. basis. We use a min-max normalization and log-scaling on all feature vectors to enhance the predictive quality by ensuring that the features are distributed evenly. The predicted self-energy is symmetrized with respect to the diagonal axis.

For a given feature vector $\Tilde{\mathbf{f}}$, our KRR model \hbox{$f^m = f^{\text{on/off},\text{Re/Im}, \omega_k}$} predicts: 
\begin{equation}\label{eq:KRRprediction}
    f^{m}(\Tilde{\mathbf{f}}) = \sum_{n = 1}^{N} \alpha^{m}_{n} K(\Tilde{\mathbf{f}},\mathbf{f}_{n}) \,,
\end{equation}
where $K$ is the kernel function. We use a superposition of three ($K_3$) Matérn covariance functions\cite{10.7551/mitpress/3206.001.0001, 10.5555/944790.944815}:
\begin{equation}\label{eq:matern3}
    K_3 = 1 + M_1 + 0.5M_{0.5} + 0.25M_{0.1}
\end{equation}
or, if mentioned, six ($K_6$) kernel functions:
\begin{equation}\label{eq:matern6}
    K_6 = K_3 + 0.4 M_{0.35} + 0.33 M_{0.25} + 0.1 M_{0.05} \,.
\end{equation}
The individual Matérn kernels $M_l = M_{l}(\mathbf{x},\mathbf{y})$ are defined as 
\begin{equation}
    M_l(\mathbf{x},\mathbf{y})_{} = \frac{1}{\Gamma(\nu)2^{\nu-1}}\Bigg(
\frac{\sqrt{2\nu}}{l} 
\|\mathbf{x},\mathbf{y}\|
\Bigg)^\nu I_\nu\Bigg(
\frac{\sqrt{2\nu}}{l} 
\|\mathbf{x},\mathbf{y}\|
\Bigg) \,,
\end{equation}
where $\|\mathbf{x},\mathbf{y}\|$ is the Euclidean distance between two vectors $\mathbf{x}$ and $\mathbf{y}$, $l$ is the length scale, $\Gamma(\nu)$ is the Gamma function, and $I_\nu(z)$ is the modified Bessel function\cite{10.7551/mitpress/3206.001.0001}. We always use $\nu = 1.5$ and only vary $l$\cite{scikit-learn}. 
The prediction in eq. \ref{eq:KRRprediction} depends on the distance defined by $K$ between our input feature vector $\Tilde{\mathbf{f}}$ and the feature vector of the training data $\mathbf{f}_n$ for $n=1,...,N$, where $N = N_d \cdot N_{\text{on/off}}$ denotes the number of training data points, which depend on the number of on/off-diagonal elements $N_{\text{on/off}}$ and the number of FCI-GF provided for training $N_d$. The regression coefficients $\boldsymbol{\alpha}^m = [\alpha^m_1, ..., \alpha^m_N]^T$ are determined by inverting the kernel matrix $\mathbf{K}$ with a small regularization parameter $\lambda = 10^{-4}$:
\begin{equation}
    \boldsymbol{\alpha}^{m} = (\mathbf{K} + \lambda \mathbf{1})^{-1} \mathbf{t}^{m} \,,
\end{equation}
where $\mathbf{t}^{m}$ is a vector of self-energy targets from the training data, see eq. \ref{eq:learning}, and $\mathbf{K}$ is defined by its entries $\mathbf{K}_{nm} = K(\mathbf{f}_n,\mathbf{f}_m)$.
This is equivalent to the following loss function
\begin{equation}
    \mathcal{L}(\boldsymbol{\alpha}^m) = \left \Vert \mathbf{K} \boldsymbol{\alpha}^m  - \mathbf{t}^m\right \Vert^{2} + \lambda (\boldsymbol{\alpha}^m)^T \mathbf{K} \boldsymbol{\alpha}^m \,,
\end{equation}
which depends on the residuals between predictions $\mathbf{K} \boldsymbol{\alpha}^m$ and targets $\mathbf{t}^m$\cite{maxwellingkrr}. We use scikit-learn in Python for the ML workflow\cite{scikit-learn}.

\subsection{Metrics of Accuracy}
We use the absolute relative difference (ARD)\cite{PhysRevB.90.155136} between the self-energy prediction $\Sigma^{\text{pred}}$ and the exact solution $\Sigma^{\text{exact}}$ to evaluate the accuracy of our ML model. This metric represents a percentage deviation from the exact result. The ARD can be defined as a function of $U$-values in the test or training data
\begin{equation}\label{eq:difference1}
       \text{ARD}(U) \vert_{\text{on},\text{off}} = \frac{1}{N_{\text{on/off}} N_{\omega_k}} \sum_{i,j} \sum_{k = 1}^{N_{\omega_k}} \frac{\abs{\Sigma_{i,j,k}^{\text{pred}} - \Sigma_{i,j,k}^{\text{exact}}}}{\abs{\Sigma_{i,j,k}^{\text{exact}}}} 
\end{equation}
which we calculate separately for all $N_{\text{on}} = L$ on-diagonal ($i=j$)  and $N_{\text{off}} = L^2 - L$  off-diagonal ($i\neq j$) matrix elements, respectively, as an average over all frequencies $k = 1, ..., N_{\omega_k}$. 
Another quantity of interest is the ARD as an average over all matrix elements for a fixed frequency:
\begin{equation}\label{eq:difference2}
    \text{ARD}(U) \vert_{\omega_n} = \frac{1}{N_{\text{on}} + N_{\text{off}}} \sum_{i,j} \frac{\abs{\Sigma_{i,j,k}^{\text{pred}} - \Sigma_{i,j,k}^{\text{exact}}}}{\abs{\Sigma_{i,j,k}^{\text{exact}}}} \,.
\end{equation}
Similarly, we can define the absolute relative difference as a function of imaginary frequencies $\text{ARD}(\omega_n)$ for all on- and off-diagonal elements ($\text{ARD}(\omega_n) \vert_{\text{on},\text{off}}$) and for all $U$-values ($\text{ARD}(\omega_n) \vert_{U}$). 

\section{Results and discussion}\label{sec:results}
\subsection{Nearest-Neighbor Interactions}\label{sec:nn}
\subsubsection{Test Data}
\begin{figure*}[htbp!]
 \centering
 \begin{adjustbox}{center}
   \includegraphics[width=2.05\columnwidth]{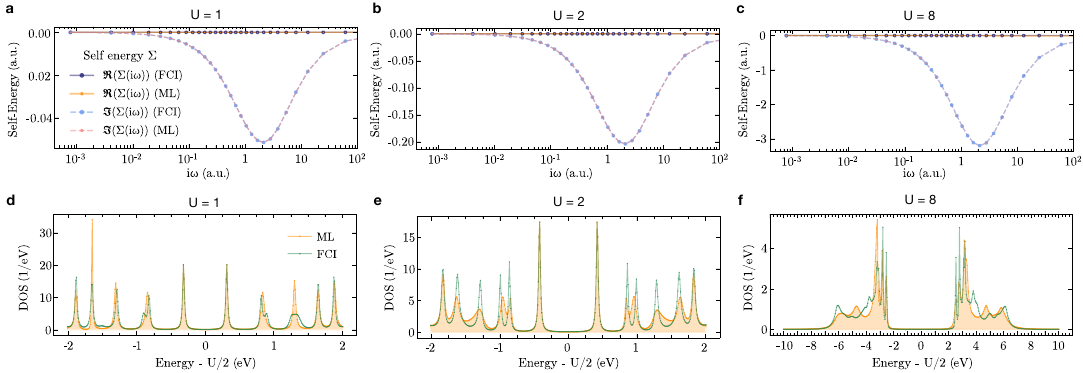}
 \end{adjustbox}
    \caption{Self-energy and DOS for unseen test data of a Hubbard model with nearest-neighbor interactions $t=1$. (a)-(c) Local self-energy on leftmost site $\Sigma_{1,1}(\mathrm{i}\omega)$ versus frequency for $U = 1$, $U = 2$, and $U = 8$, respectively. Machine-learned prediction (ML, orange, small circles) versus exact solution (FCI, blue, large circles). Solid lines: real part. Dashed lines: imaginary part. (d)-(f) DOS from FCI (green line) compared to ML prediction (orange filled area) around the Fermi energy $E_F = U/2$.}
 \label{fig:fig1}
\end{figure*}

First, we investigate the predictive power of our ML framework in the case of nearest-neighbor interactions. 
In fig. \ref{fig:fig1}(a)-(c), we present the local self-energy $\Sigma_{i,i}$ on the first site $i=1$ as a function of imaginary frequencies for different Coulomb interaction strengths $U=1$, $U=2$, and $U=8$, respectively. All of these values correspond to test data not used during training. We observe indistinguishable curves for the predictions (ML, orange, small circles) in comparison to the exact function (FCI, blue, large circles), both for the imaginary part (dashed lines) and the near-zero real part (solid lines). In fig. \ref{fig:fig1}(d)-(f), we compare the exact density of states (DOS) obtained from FCI with the ML prediction for the same $U$-values presented in the top row. The ML model predicts the energies of all peaks and the approximate shape and height correctly, although the precision is reduced further away from the Fermi edge. The predicted DOS matches the exact DOS better when the system is weakly correlated, compare, e.g., fig. \ref{fig:fig1}(d) to (f). However, it is important to note that the DOS is plotted over a larger range of energies to show the full electronic structure.

In the following, we present both the DOS and the self-energy, but it is important to note that our ML model only predicts the Green's function on the imaginary axis, while the DOS is obtained via analytic continuation (A.C.), which introduces additional errors. The FCI-GF is computed on the imaginary axis, ensuring that the self-energy target is exact. This is also a computationally less expensive way of obtaining large amounts of training data. However, the true FCI-GF computed directly on real frequencies may deviate from the reference function to which we compare our ML predictions. Hence, one may not accurately assess the quality of this model by evaluating the DOS, which is why we use the absolute relative difference (ARD) of the self-energy, see eq. \ref{eq:difference1} and \ref{eq:difference2}, to estimate the accuracy of our ML model. We argue that the main source of error in the predicted DOS originates from the A.C. when transforming the Green's function to the real axis, not from the quality of the ML predictions on the imaginary axis. To support this, we compare the results for test data to the DOS for training data, and we present a comprehensive overview of the ARD.

\subsubsection{Training Data}
\begin{figure*}[htbp!]
 \centering
 \begin{adjustbox}{center}
   \includegraphics[width=2.05\columnwidth]{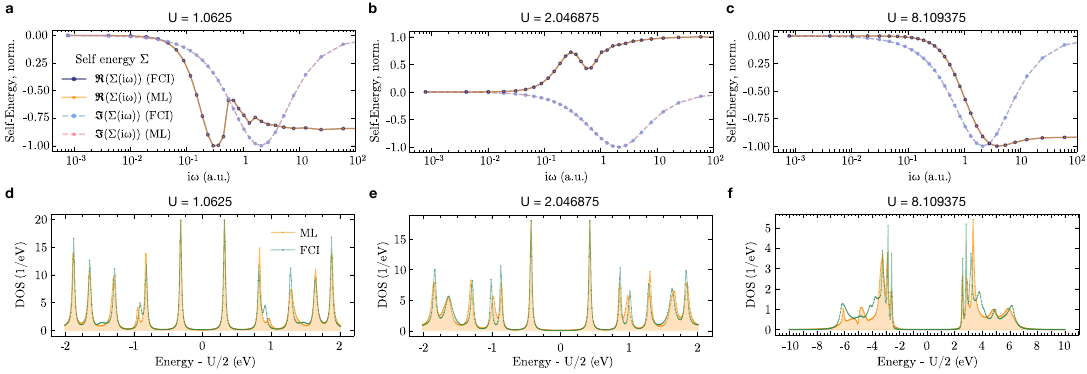}
 \end{adjustbox}
    \caption{Same as fig. \ref{fig:fig1}, but for training data. (a)-(c) Local self-energy $\Sigma_{1,1}(\mathrm{i}\omega)$ in normalized units versus frequency for $U = 1.0625$, $U = 2.046875$, and $U = 8.109375$, respectively, and (d)-(f) DOS around the Fermi energy for the same $U$-values.}
 \label{fig:fig2}
\end{figure*}

In fig. \ref{fig:fig2}(a)-(c), we show the local self-energy on the first site 
for training data with $U=1.0625$, $U=2.046875$, and $U=8.109375$, respectively. Note that these data are presented in normalized units. The ML prediction and the exact solution are visually indistinguishable on the scale of the plot. Similar to the previous results, we observe that the ML model accurately reproduces the peak shapes and heights in the DOS, see fig. \ref{fig:fig2}(d)-(f). We observe similar deviations between the predicted DOS and the exact result when comparing test data in fig. \ref{fig:fig1}(f) and training data in fig. \ref{fig:fig2}(f). This supports our claim that the main source of deviations stems from the A.C. method, which becomes clear when we compare the self-energy ARD between test and training data. In the supplementary material, we present additional data to quantify the DOS error, including a comparison of the Padé A.C. to a method based on the pole-expansion scheme (PES)\cite{PhysRevB.107.075151}.

\subsubsection{Accuracy of Predictions}
\begin{figure*}[htbp!]
 \centering
 \begin{adjustbox}{center}
   \includegraphics[width=2.05\columnwidth]{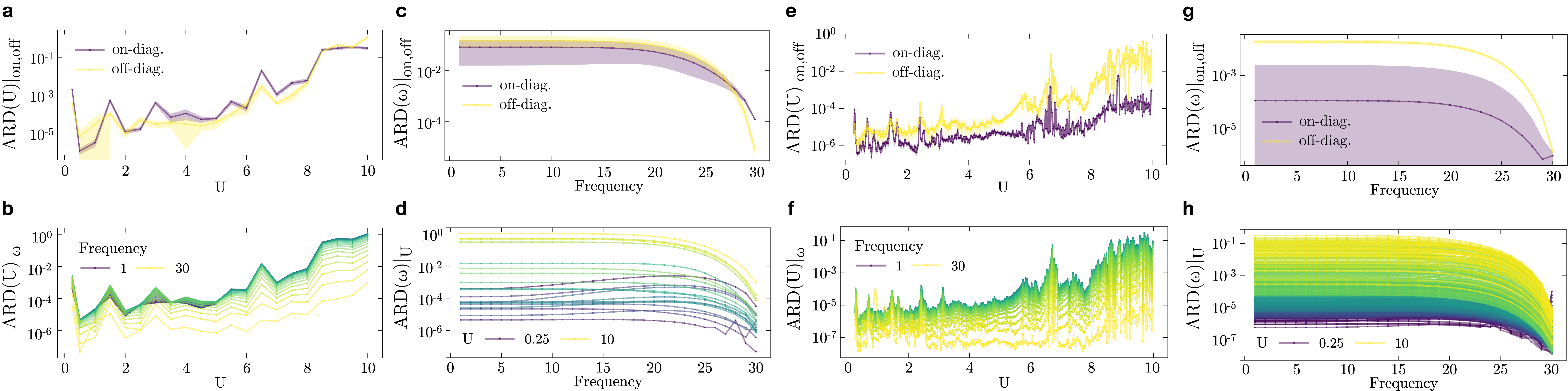}
 \end{adjustbox}
    \caption{Overview of absolute relative difference (ARD) between exact and ML-predicted self-energy of the Hubbard model with nearest-neighbor hopping $t=1$. (a)-(d) show results for test data, and (e)-(h) for training data. (a) ARD(U)$\vert_{\text{on,off}}$ for all on- and off-diagonal matrix elements, showing that systems with large $U$ can be predicted less accurately. (b) ARD(U)$\vert_{\omega}$ for individual frequencies from smallest $\omega_1$ (purple) to largest $\omega_{30}$ (yellow) shows reduced accuracy at small frequencies. (c) ARD($\omega$)$\vert_{\text{on,off}}$ as a function of frequency over all $U$ shows nearly constant error across most frequencies. 
    (d) ARD($\omega$)$\vert_{U}$ for different interaction strengths from $U=0.25$ (purple) to $U = 10$ (yellow) as a function of frequency index. Note that no error bars are plotted for the bottom plots. (e)–(h) Corresponding quantities for the training data.}
 \label{fig:fig3}
\end{figure*}

The absolute relative difference (ARD) between the exact and predicted self-energy for test data from fig. \ref{fig:fig1} is presented in fig. \ref{fig:fig3}(a)-(d). Figure \ref{fig:fig3}(a) shows the ARD for on-diagonal (purple) and off-diagonal (yellow) matrix elements over all frequencies as a function of Coulomb repulsion $U$, see eq. \ref{eq:difference1}. The colored areas correspond to the standard deviation around the mean computed in linear space. When the frequency-resolved ARD is displayed on a logarithmic axis, the errors may appear visually large, due to the broad distribution of ARD values. The predictions have the lowest error for small and intermediate $U$-values, with relative differences between $10^{-5}$ and $10^{-4}$. In contrast, systems with stronger correlations ($U>8$) can be predicted less accurately, leading to a relative error of up to 10\%. Interestingly, there is a large range of Coulomb repulsions up to $U \approx 6$ that can be predicted with the same relative error. This raises the question of whether the increased error at large $U$ arises from complex excitation properties that are difficult to capture with mean-field data, and whether it can be mitigated by including training samples at even larger Coulomb interactions. In fig. \ref{fig:fig3}(b), we present the ARD over all matrix elements for different frequencies. Similar to fig. \ref{fig:fig3}(a), we observe that the error increases with increasing Coulomb repulsion $U$. We use an adapted Gauss-Legendre grid, meaning that the mapping of frequency index to frequency value is not linear (see fig. \ref{fig:fig1}(d)-(f)), where $\omega_{1}$ (purple) represents the smallest frequency value, and $\omega_{30}$ (yellow) the largest value. We observe that the smallest frequency values (purple) can be predicted less accurately due to the more complicated low-energy physics near the Fermi edge. 
Each data point in fig. \ref{fig:fig3}(c) stems from an independent ML model, see eq. \ref{eq:difference2}, which shows that our ML architecture is well-suited to learn the self-energy over a broad range of frequencies. Similar to fig. \ref{fig:fig3}(b), we observe that the self-energy of the largest frequency values can be predicted with errors as low as $10^{-4}$, while the relative error increases for smaller frequencies.
In fig. \ref{fig:fig3}(d), we present the ARD over all matrix elements for different $U$-values as a function of frequency. 
We find that the best predictions can be achieved for small and intermediate $U$-values, but not for the smallest value $U = 0.25$. This is most likely due to the fact that we did not provide training data for $U<0.25$. 

We present the same overview of the ARD for our training data in fig. \ref{fig:fig3}(e)-(h). Similar to test data, we find that the ARD is lowest for small and intermediate $U$-values up to $U \approx 6$, see fig. \ref{fig:fig3}(e), while the relative error increases for larger $U$-values. The on-diagonal elements are predicted with an ARD of $10^{-4}$ for the largest $U$-values, while we find relative errors as low as $10^{-6}$ for smaller $U$-values. Note that the diagonal elements of the self-energy are more important for the predictions than the off-diagonal ones. We observe lower relative errors for training data compared to test data, compare fig. \ref{fig:fig3}(f) to (b), respectively. This increases our confidence in the model's ability to memorize the training data, meaning that the mean-field features are sufficient for our KRR model to reproduce many-body spectral quantities. In the appendix, we show that we are not providing the self-energy target in the feature vector by including a first-order $G_0 W_0$ calculation. The observation that both fig. \ref{fig:fig3}(e) and (f) have similar shapes when comparing on- versus off-diagonal elements over different $U$-values or frequencies, respectively, shows that our ML model is a sensible approach to learn the full self-energy function, including on-diagonal elements representing local interaction corrections and off-diagonal non-local correlation effects between different sites. This can also be inferred from the relatively constant ARD over a large range of frequencies in fig. \ref{fig:fig3}(g) and (h). 

For models with nearest-neighbor hopping, the DOS as a function of energy is symmetric w.r.t. the Fermi level at $U/2$, which could be used to enhance the accuracy of the A.C., for example, by only predicting a subset of the energies. 
Another way of improving the results is by increasing the amount of training data, which we expect\cite{PhysRevB.90.155136} will lead to better results, a direction we did not pursue due to the computational cost of the FCI calculations.

\subsection{Long-Range Interactions}\label{sec:longrange}
\subsubsection{Test Data}
\begin{figure}[htbp!]
 \centering
 \begin{adjustbox}{center}
   \includegraphics[width=1.025\columnwidth]{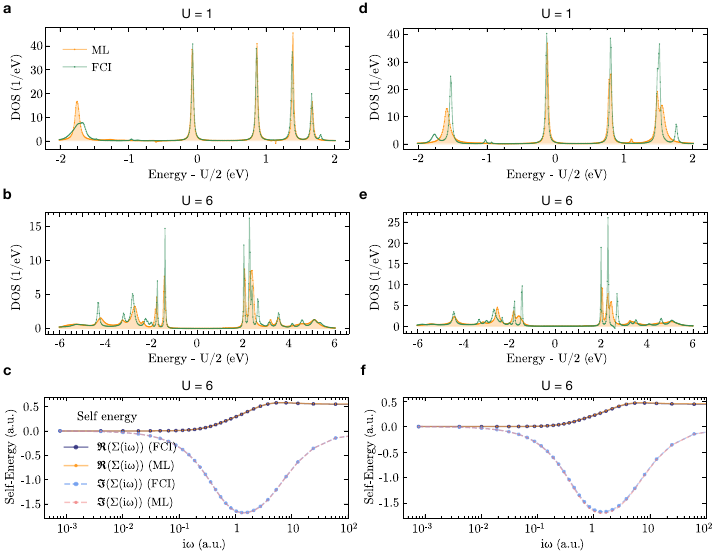}
 \end{adjustbox}
    \caption{Test data for a Hubbard model with $t = 1$, $t' = 0.25$, and $t''=0.1$. Comparison of ML prediction to FCI for (a) and (b) DOS at $U = 1$ and $U = 6$, respectively, and (c) local self-energy $\Sigma_{1,1}(\mathrm{i}\omega)$ versus frequency for $U=6$. (d) and (e) DOS for a system with additional interaction $t''' = 0.1$ for $U = 1$ and $U = 6$, respectively, and (f) local self-energy for $U=6$.}
 \label{fig:fig4}
\end{figure}

A current challenge with most computational methods is capturing long-range interactions. In the following, we focus on a Hubbard model with nearest-neighbor hopping $t=1$ and long-range hopping terms $t' = 0.25$ and $t''=0.1$. In fig. \ref{fig:fig4}(a) and (b), we present the exact and predicted DOS for unseen test data with $U=1$ and $U=6$, respectively. Note that  the Fermi level is shifted from $U/2$. The predicted DOS is close to the exact result, and the ML model captures both the position and relative height of most peaks. The predicted local self-energy on the first site shows slight deviations from the exact solution for frequencies around $\mathrm{i}\omega \approx 1$ for $U = 6$, see fig. \ref{fig:fig4}(c).
The same data is presented in fig. \ref{fig:fig4}(d)-(f) for a system with an additional hopping term $t''' = 0.1$. Note that we use a kernel with six ($K_6$) instead of three ($K_3$) Matérn functions for calculations with $t''' \neq 0$. We provide a comprehensive comparison between these kernels in the appendix\cite{supp}. Our improved kernel $K_6$ is able to capture the relationship between features and targets more accurately compared to $K_3$ for the same system parameters, increasing the accuracy for the case with $t''' \neq 0$.

\begin{figure*}[htbp!]
 \centering
 \begin{adjustbox}{center}
   \includegraphics[width=2.05\columnwidth]{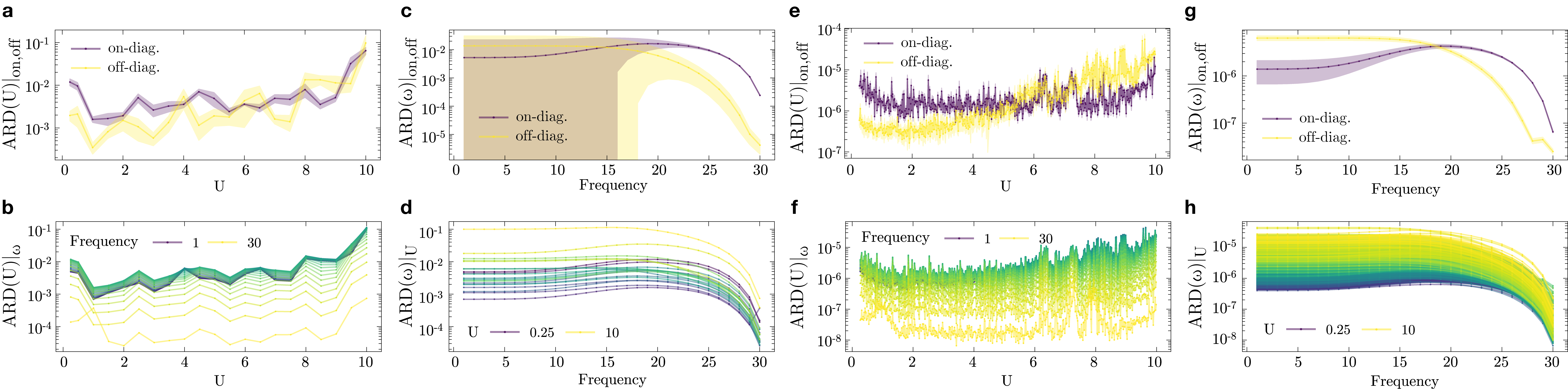}
 \end{adjustbox}
    \caption{Same as fig. \ref{fig:fig3}, but for a Hubbard model with long-range interactions $t = 1$, $t' = 0.25$, and $t''=0.1$. ARD for (a)-(d) test data and (e)-(h) training data.}
 \label{fig:fig5}
\end{figure*}

\subsubsection{Accuracy of Predictions}
In fig. \ref{fig:fig5}, we provide an overview of the ARD for a system with long-range hopping terms $t'=0.25$, $t''=0.1$, and $t'''=0$.
The ARD increases for test data with larger $U$-values, see fig. \ref{fig:fig5}(a), although less pronounced compared to the trends in fig. \ref{fig:fig3}(a). We achieve relative errors between $10^{-4}$ and $10^{-2}$ over a large range of interaction strengths (fig. \ref{fig:fig5}(b)), where the best accuracy is reached at low and intermediate $U$-values, see also fig. \ref{fig:fig5}(d). The ARD curve in fig. \ref{fig:fig5}(c) is different from the case with purely nearest-neighbor interactions in fig. \ref{fig:fig3}(c), which can be explained by a different frequency dependence of $\Sigma(\mathrm{i}\omega)$ for systems with long-range interactions. Similarly, we see in fig. \ref{fig:fig5}(d) that the largest ARD occurs at larger frequency values around $\omega_{20}$ instead of $\omega_1$. In fig. \ref{fig:fig5}(e)-(h), we present the same metrics for our training data. The ARD for the off-diagonal matrix elements (yellow) increases with the on-site Coulomb repulsion $U$ (fig. \ref{fig:fig5}(e)), because off-diagonal elements corresponding to long-range interactions become more important for systems with larger $U$. On the other hand, the ARD does not show this trend for on-diagonal terms (purple).
Long-range hopping terms broaden the underlying dispersion and weaken sharp momentum-space features, leading to a self-energy that varies more smoothly in momentum space across different interaction strengths compared to the nearest-neighbor case presented in fig. \ref{fig:fig3}. As a result, the relationship between the mean-field features and self-energy targets learned by the ML model remains structurally similar for different U, yielding a weak $U$-dependence of the ARD.
This highlights why training separate models for on- and off-diagonal elements is a useful design choice. For the training data, the relative differences are as low as $10^{-5}$ and $10^{-8}$ for small and large frequencies, respectively, indicating near-exact results (fig. \ref{fig:fig5}(f)). This increases our confidence that this ML model can relate mean-field features to the many-body self-energy. We observe that the ARD is less dependent on $U$ for systems with long-range interactions, see fig. \ref{fig:fig5}(f) in comparison to fig. \ref{fig:fig3}(f). This is another indication that our method might be particularly well suited for systems with complicated long-range interactions that modify the self-energy function in a favorable way for our ML framework, see also fig. \ref{fig:fig5}(g) in comparison to the nearest-neighbor Hubbard model in fig. \ref{fig:fig3}(g). 

In the appendix, we present the same overview for the case with $t''' = 0.1$\cite{supp}. Similar to the previous results, we find an ARD as low as $10^{-3}$ and $10^{-7}$ over a large range of $U$-values for test and training data, respectively.
\section{Conclusion} \label{sec:conclusion}
In this work, we demonstrate that an ML framework based on Kernel Ridge Regression (KRR) can learn and predict the imaginary-frequency self-energy function of one-dimensional Hubbard models across a wide range of on-site interaction strengths, including both short- and long-range hopping terms. In all examples, the predicted self-energies exhibit a low absolute relative difference (ARD) compared to the exact results. Using Padé analytic continuation (A.C.), we transform the self-energy calculated on the imaginary frequency axis to the real energy axis, enabling us to predict the density of states (DOS). We attribute the main source of error in the DOS predictions to the A.C. method rather than to the ML model. Systems with long-range hopping terms are predicted with relative errors comparable to those of a nearest-neighbor model, suggesting that our method remains robust for real materials with complex interactions.
Our findings indicate that the mean-field quantities of the one-dimensional Hubbard model contain sufficient information for a KRR model to reproduce and predict many-body self-energies and capture the signatures of strong correlations in spectral observables. We show that the highly nonlinear structure of deep neural networks, which makes them challenging to interpret, is not required to capture the spectral features of strongly correlated quantum systems. Instead, our results suggest that simpler ML models such as KRR might be sufficient to learn signatures of strong correlations.

In future work, our ML framework could be extended to predict spectral properties of lattice models with varying electron and hole doping, system sizes, dimensionality, or interaction strengths, possibly by employing a Mixture-of-Experts (MoE) approach.
While our ML architecture does not explicitly encode system size, extending the approach to the thermodynamic limit remains an unresolved challenge. The success of our ML model relies on the relationship between mean-field features at the Hartree-Fock level and the imaginary-frequency self-energy function, both of which are generally less sensitive to system size than real-frequency spectral features\cite{PhysRevB.106.235106,RevModPhys.77.1027,doi.org/10.1140/epjs/s11734-023-00976-5}. On this basis, we believe that the present framework could, in principle, be extended to predict the imaginary-frequency self-energy for systems of different sizes.
Therefore, a natural next step is to investigate how well our model generalizes to larger systems or to related systems such as the \hbox{$t-J$-model}.
Another open question concerns the connection between one- and two-dimensional lattice models. Our framework may help explore whether two-dimensional correlations can be inferred from one-dimensional systems, possibly advancing the understanding of topological phenomena\cite{PhysRevX.8.021048} or superconductivity\cite{PhysRevB.71.104520}.

\section*{Acknowledgements} \label{sec:acknowledgements}
I am grateful for the support from Tianyu Zhu, as well as for fruitful discussions and the hospitality of his group. I would like to thank Uli Schollwöck, Chris Roth, Jason Kaye, and Andy Millis for their helpful comments on this manuscript. 
I thank Christian Venturella, Christopher Hillenbrand, and Jiachen Li for helpful discussions and support with the MLGF library. 
I am grateful for the hospitality of the Center for Computational Quantum Physics at the Flatiron Institute. The Flatiron Institute is a division of the Simons Foundation. I acknowledge funding from the German Academic Scholarship Foundation and the Swiss Study Foundation. 
\bibliography{ref}

@article{doi:10.1021/acs.jctc.3c01146,
	author = {Venturella, Christian and Hillenbrand, Christopher and Li, Jiachen and Zhu, Tianyu},
	doi = {10.1021/acs.jctc.3c01146},
	eprint = {https://doi.org/10.1021/acs.jctc.3c01146},
	journal = {Journal of Chemical Theory and Computation},
	note = {PMID: 38150268},
	number = {1},
	pages = {143-154},
	title = {Machine Learning Many-Body Green's Functions for Molecular Excitation Spectra},
	url = {https://doi.org/10.1021/acs.jctc.3c01146},
	volume = {20},
	year = {2024}}

@article{PhysRevX.11.021006,
  title = {Ab Initio Full Cell $GW+\mathrm{DMFT}$ for Correlated Materials},
  author = {Zhu, Tianyu and Chan, Garnet Kin-Lic},
  journal = {Phys. Rev. X},
  volume = {11},
  issue = {2},
  pages = {021006},
  numpages = {13},
  year = {2021},
  month = {Apr},
  publisher = {American Physical Society},
  doi = {10.1103/PhysRevX.11.021006},
  url = {https://link.aps.org/doi/10.1103/PhysRevX.11.021006}
}

@article{10.1021/acs.jctc.9b00934,
	author = {Zhu, Tianyu and Cui, Zhi-Hao and Chan, Garnet Kin-Lic},
	date = {2020/01/14},
	doi = {10.1021/acs.jctc.9b00934},
	isbn = {1549-9618},
	journal = {Journal of Chemical Theory and Computation},
	journal1 = {Journal of Chemical Theory and Computation},
	journal2 = {J. Chem. Theory Comput.},
	month = {01},
	number = {1},
	pages = {141--153},
	publisher = {American Chemical Society},
	title = {Efficient Formulation of Ab Initio Quantum Embedding in Periodic Systems: Dynamical Mean-Field Theory},
	type = {doi: 10.1021/acs.jctc.9b00934},
	url = {https://doi.org/10.1021/acs.jctc.9b00934},
	volume = {16},
	year = {2020},
	year1 = {2020}}

@article{PhysRevB.100.115154,
  title = {Coupled-cluster impurity solvers for dynamical mean-field theory},
  author = {Zhu, Tianyu and Jim\'enez-Hoyos, Carlos A. and McClain, James and Berkelbach, Timothy C. and Chan, Garnet Kin-Lic},
  journal = {Phys. Rev. B},
  volume = {100},
  issue = {11},
  pages = {115154},
  numpages = {9},
  year = {2019},
  month = {Sep},
  publisher = {American Physical Society},
  doi = {10.1103/PhysRevB.100.115154},
  url = {https://link.aps.org/doi/10.1103/PhysRevB.100.115154}
}

@article{doi:10.1021/acs.jctc.0c00704,
	author = {Zhu, Tianyu and Chan, Garnet Kin-Lic},
	doi = {10.1021/acs.jctc.0c00704},
	eprint = {https://doi.org/10.1021/acs.jctc.0c00704},
	journal = {Journal of Chemical Theory and Computation},
	note = {PMID: 33397095},
	number = {2},
	pages = {727-741},
	title = {All-Electron Gaussian-Based G0W0 for Valence and Core Excitation Energies of Periodic Systems},
	url = {https://doi.org/10.1021/acs.jctc.0c00704},
	volume = {17},
	year = {2021}}

@article{10.1063/5.0006074,
	author = {Sun, Qiming and Zhang, Xing and Banerjee, Samragni and Bao, Peng and Barbry, Marc and Blunt, Nick S. and Bogdanov, Nikolay A. and Booth, George H. and Chen, Jia and Cui, Zhi-Hao and Eriksen, Janus J. and Gao, Yang and Guo, Sheng and Hermann, Jan and Hermes, Matthew R. and Koh, Kevin and Koval, Peter and Lehtola, Susi and Li, Zhendong and Liu, Junzi and Mardirossian, Narbe and McClain, James D. and Motta, Mario and Mussard, Bastien and Pham, Hung Q. and Pulkin, Artem and Purwanto, Wirawan and Robinson, Paul J. and Ronca, Enrico and Sayfutyarova, Elvira R. and Scheurer, Maximilian and Schurkus, Henry F. and Smith, James E. T. and Sun, Chong and Sun, Shi-Ning and Upadhyay, Shiv and Wagner, Lucas K. and Wang, Xiao and White, Alec and Whitfield, James Daniel and Williamson, Mark J. and Wouters, Sebastian and Yang, Jun and Yu, Jason M. and Zhu, Tianyu and Berkelbach, Timothy C. and Sharma, Sandeep and Sokolov, Alexander Yu. and Chan, Garnet Kin-Lic},
	doi = {10.1063/5.0006074},
	eprint = {https://pubs.aip.org/aip/jcp/article-pdf/doi/10.1063/5.0006074/16722275/024109\_1\_online.pdf},
	issn = {0021-9606},
	journal = {The Journal of Chemical Physics},
	month = {07},
	number = {2},
	pages = {024109},
	title = {Recent developments in the PySCF program package},
	url = {https://doi.org/10.1063/5.0006074},
	volume = {153},
	year = {2020}}

@article{https://doi.org/10.1002/wcms.1340,
	author = {Sun, Qiming and Berkelbach, Timothy C. and Blunt, Nick S. and Booth, George H. and Guo, Sheng and Li, Zhendong and Liu, Junzi and McClain, James D. and Sayfutyarova, Elvira R. and Sharma, Sandeep and Wouters, Sebastian and Chan, Garnet Kin-Lic},
	doi = {https://doi.org/10.1002/wcms.1340},
	eprint = {https://wires.onlinelibrary.wiley.com/doi/pdf/10.1002/wcms.1340},
	journal = {WIREs Computational Molecular Science},
	number = {1},
	pages = {e1340},
	title = {PySCF: the Python-based simulations of chemistry framework},
	url = {https://wires.onlinelibrary.wiley.com/doi/abs/10.1002/wcms.1340},
	volume = {8},
	year = {2018}}

@article{PhysRevLett.69.2863,
  title = {Density matrix formulation for quantum renormalization groups},
  author = {White, Steven R.},
  journal = {Phys. Rev. Lett.},
  volume = {69},
  issue = {19},
  pages = {2863--2866},
  numpages = {0},
  year = {1992},
  month = {Nov},
  publisher = {American Physical Society},
  doi = {10.1103/PhysRevLett.69.2863},
  url = {https://link.aps.org/doi/10.1103/PhysRevLett.69.2863}
}

@article{SCHOLLWOCK201196,
	author = {Ulrich Schollw{\"o}ck},
	doi = {https://doi.org/10.1016/j.aop.2010.09.012},
	issn = {0003-4916},
	journal = {Annals of Physics},
	note = {January 2011 Special Issue},
	number = {1},
	pages = {96-192},
	title = {The density-matrix renormalization group in the age of matrix product states},
	url = {https://www.sciencedirect.com/science/article/pii/S0003491610001752},
	volume = {326},
	year = {2011}}

@article{PhysRevB.80.075116,
  title = {Quantum Monte Carlo study of the two-dimensional fermion Hubbard model},
  author = {Varney, C. N. and Lee, C.-R. and Bai, Z. J. and Chiesa, S. and Jarrell, M. and Scalettar, R. T.},
  journal = {Phys. Rev. B},
  volume = {80},
  issue = {7},
  pages = {075116},
  numpages = {8},
  year = {2009},
  month = {Aug},
  publisher = {American Physical Society},
  doi = {10.1103/PhysRevB.80.075116},
  url = {https://link.aps.org/doi/10.1103/PhysRevB.80.075116}
}

@article{doi:10.1126/science.aag2302,
	author = {Giuseppe Carleo and Matthias Troyer},
	doi = {10.1126/science.aag2302},
	eprint = {https://www.science.org/doi/pdf/10.1126/science.aag2302},
	journal = {Science},
	number = {6325},
	pages = {602-606},
	title = {Solving the quantum many-body problem with artificial neural networks},
	url = {https://www.science.org/doi/abs/10.1126/science.aag2302},
	volume = {355},
	year = {2017}}

@article{doi.org/10.1038/s41467-020-15724-9,
	author = {Choo, Kenny and Mezzacapo, Antonio and Carleo, Giuseppe},
	date = {2020/05/12},
	doi = {10.1038/s41467-020-15724-9},
	id = {Choo2020},
	isbn = {2041-1723},
	journal = {Nature Communications},
	number = {1},
	pages = {2368},
	title = {Fermionic neural-network states for ab-initio electronic structure},
	url = {https://doi.org/10.1038/s41467-020-15724-9},
	volume = {11},
	year = {2020}}

@article{PhysRevLett.121.167204,
  title = {Symmetries and Many-Body Excitations with Neural-Network Quantum States},
  author = {Choo, Kenny and Carleo, Giuseppe and Regnault, Nicolas and Neupert, Titus},
  journal = {Phys. Rev. Lett.},
  volume = {121},
  issue = {16},
  pages = {167204},
  numpages = {6},
  year = {2018},
  month = {Oct},
  publisher = {American Physical Society},
  doi = {10.1103/PhysRevLett.121.167204},
  url = {https://link.aps.org/doi/10.1103/PhysRevLett.121.167204}
}

@article{annurev:/content/journals/10.1146/annurev-conmatphys-020911-125018,
	author = {Stoudenmire, E.M. and White, Steven R.},
	doi = {https://doi.org/10.1146/annurev-conmatphys-020911-125018},
	issn = {1947-5462},
	journal = {Annual Review of Condensed Matter Physics},
	number = {Volume 3, 2012},
	pages = {111-128},
	publisher = {Annual Reviews},
	title = {Studying Two-Dimensional Systems with the Density Matrix Renormalization Group},
	type = {Journal Article},
	url = {https://www.annualreviews.org/content/journals/10.1146/annurev-conmatphys-020911-125018},
	volume = {3},
	year = {2012}}

@article{PhysRevB.41.9301,
  title = {Sign problem in the numerical simulation of many-electron systems},
  author = {Loh, E. Y. and Gubernatis, J. E. and Scalettar, R. T. and White, S. R. and Scalapino, D. J. and Sugar, R. L.},
  journal = {Phys. Rev. B},
  volume = {41},
  issue = {13},
  pages = {9301--9307},
  numpages = {0},
  year = {1990},
  month = {May},
  publisher = {American Physical Society},
  doi = {10.1103/PhysRevB.41.9301},
  url = {https://link.aps.org/doi/10.1103/PhysRevB.41.9301}
}

@article{https://doi.org/10.1002/qua.24939,
	author = {Vu, Kevin and Snyder, John C. and Li, Li and Rupp, Matthias and Chen, Brandon F. and Khelif, Tarek and M{\"u}ller, Klaus-Robert and Burke, Kieron},
	doi = {https://doi.org/10.1002/qua.24939},
	eprint = {https://onlinelibrary.wiley.com/doi/pdf/10.1002/qua.24939},
	journal = {International Journal of Quantum Chemistry},
	number = {16},
	pages = {1115-1128},
	title = {Understanding kernel ridge regression: Common behaviors from simple functions to density functionals},
	url = {https://onlinelibrary.wiley.com/doi/abs/10.1002/qua.24939},
	volume = {115},
	year = {2015}}

@book{fetter2003quantum,
  title={Quantum Theory of Many-particle Systems},
  author={Fetter, A.L. and Walecka, J.D.},
  isbn={9780486428277},
  lccn={2003043536},
  series={Dover Books on Physics},
  url={https://books.google.com/books?id=0wekf1s83b0C},
  year={2003},
  publisher={Dover Publications}
}

@article{PhysRevResearch.6.043280,
  title = {Fine-tuning neural network quantum states},
  author = {Rende, Riccardo and Goldt, Sebastian and Becca, Federico and Viteritti, Luciano Loris},
  journal = {Phys. Rev. Res.},
  volume = {6},
  issue = {4},
  pages = {043280},
  numpages = {9},
  year = {2024},
  month = {Dec},
  publisher = {American Physical Society},
  doi = {10.1103/PhysRevResearch.6.043280},
  url = {https://link.aps.org/doi/10.1103/PhysRevResearch.6.043280}
}

@article{PhysRevB.100.245123,
  title = {Machine learning approach to dynamical properties of quantum many-body systems},
  author = {Hendry, Douglas and Feiguin, Adrian E.},
  journal = {Phys. Rev. B},
  volume = {100},
  issue = {24},
  pages = {245123},
  numpages = {8},
  year = {2019},
  month = {Dec},
  publisher = {American Physical Society},
  doi = {10.1103/PhysRevB.100.245123},
  url = {https://link.aps.org/doi/10.1103/PhysRevB.100.245123}
}

@article{Lange_2024,
	author = {Lange, Hannah and Van de Walle, Anka and Abedinnia, Atiye and Bohrdt, Annabelle},
	doi = {10.1088/2058-9565/ad7168},
	journal = {Quantum Science and Technology},
	month = {sep},
	number = {4},
	pages = {040501},
	publisher = {IOP Publishing},
	title = {From architectures to applications: a review of neural quantum states},
	url = {https://dx.doi.org/10.1088/2058-9565/ad7168},
	volume = {9},
	year = {2024}}

@article{PhysRevB.102.205122,
  title = {Phases of two-dimensional spinless lattice fermions with first-quantized deep neural-network quantum states},
  author = {Stokes, James and Moreno, Javier Robledo and Pnevmatikakis, Eftychios A. and Carleo, Giuseppe},
  journal = {Phys. Rev. B},
  volume = {102},
  issue = {20},
  pages = {205122},
  numpages = {10},
  year = {2020},
  month = {Nov},
  publisher = {American Physical Society},
  doi = {10.1103/PhysRevB.102.205122},
  url = {https://link.aps.org/doi/10.1103/PhysRevB.102.205122}
}

@article{PhysRevB.96.205152,
  title = {Restricted Boltzmann machine learning for solving strongly correlated quantum systems},
  author = {Nomura, Yusuke and Darmawan, Andrew S. and Yamaji, Youhei and Imada, Masatoshi},
  journal = {Phys. Rev. B},
  volume = {96},
  issue = {20},
  pages = {205152},
  numpages = {8},
  year = {2017},
  month = {Nov},
  publisher = {American Physical Society},
  doi = {10.1103/PhysRevB.96.205152},
  url = {https://link.aps.org/doi/10.1103/PhysRevB.96.205152}
}

@inproceedings{10.1145/3581784.3607061, author = {Wu, Yangjun and Guo, Chu and Fan, Yi and Zhou, Pengyu and Shang, Honghui}, title = {NNQS-Transformer: an Efficient and Scalable Neural Network Quantum States Approach for Ab initio Quantum Chemistry}, year = {2023}, isbn = {9798400701092}, publisher = {Association for Computing Machinery}, address = {New York, NY, USA}, url = {https://doi.org/10.1145/3581784.3607061}, doi = {10.1145/3581784.3607061}, booktitle = {Proceedings of the International Conference for High Performance Computing, Networking, Storage and Analysis}, articleno = {42}, numpages = {13}, location = {Denver, CO, USA}, series = {SC '23} }

@book{Essler_Frahm_Göhmann_Klümper_Korepin_2005, place={Cambridge}, title={The One-Dimensional Hubbard Model}, publisher={Cambridge University Press}, author={Essler, Fabian H. L. and Frahm, Holger and Göhmann, Frank and Klümper, Andreas and Korepin, Vladimir E.}, year={2005}}

@article{doi.org/10.1007/BF02188656,
	author = {Bach, Volker and Lieb, Elliott H. and Solovej, Jan Philip},
	date = {1994/07/01},
	doi = {10.1007/BF02188656},
	id = {Bach1994},
	isbn = {1572-9613},
	journal = {Journal of Statistical Physics},
	number = {1},
	pages = {3--89},
	title = {Generalized Hartree-Fock theory and the Hubbard model},
	url = {https://doi.org/10.1007/BF02188656},
	volume = {76},
	year = {1994}}

@article{PhysRev.81.385,
  title = {A Simplification of the Hartree-Fock Method},
  author = {Slater, J. C.},
  journal = {Phys. Rev.},
  volume = {81},
  issue = {3},
  pages = {385--390},
  numpages = {0},
  year = {1951},
  month = {Feb},
  publisher = {American Physical Society},
  doi = {10.1103/PhysRev.81.385},
  url = {https://link.aps.org/doi/10.1103/PhysRev.81.385}
}

@article{PhysRev.140.A1133,
  title = {Self-Consistent Equations Including Exchange and Correlation Effects},
  author = {Kohn, W. and Sham, L. J.},
  journal = {Phys. Rev.},
  volume = {140},
  issue = {4A},
  pages = {A1133--A1138},
  numpages = {0},
  year = {1965},
  month = {Nov},
  publisher = {American Physical Society},
  doi = {10.1103/PhysRev.140.A1133},
  url = {https://link.aps.org/doi/10.1103/PhysRev.140.A1133}
}

@article{PhysRev.139.A796,
  title = {New Method for Calculating the One-Particle Green's Function with Application to the Electron-Gas Problem},
  author = {Hedin, Lars},
  journal = {Phys. Rev.},
  volume = {139},
  issue = {3A},
  pages = {A796--A823},
  numpages = {0},
  year = {1965},
  month = {Aug},
  publisher = {American Physical Society},
  doi = {10.1103/PhysRev.139.A796},
  url = {https://link.aps.org/doi/10.1103/PhysRev.139.A796}
}

@article{10.3389/fchem.2019.00377,
	author = {Golze, Dorothea and Dvorak, Marc and Rinke, Patrick},
	doi = {10.3389/fchem.2019.00377},
	issn = {2296-2646},
	journal = {Frontiers in Chemistry},
	title = {The GW Compendium: A Practical Guide to Theoretical Photoemission Spectroscopy},
	url = {https://www.frontiersin.org/journals/chemistry/articles/10.3389/fchem.2019.00377},
	volume = {Volume 7 - 2019},
	year = {2019}}

@article{https://doi.org/10.1002/qua.560440808,
	author = {Nooijen, Marcel and Snijders, Jaap G.},
	doi = {https://doi.org/10.1002/qua.560440808},
	eprint = {https://onlinelibrary.wiley.com/doi/pdf/10.1002/qua.560440808},
	journal = {International Journal of Quantum Chemistry},
	number = {S26},
	pages = {55-83},
	title = {Coupled cluster approach to the single-particle Green's function},
	url = {https://onlinelibrary.wiley.com/doi/abs/10.1002/qua.560440808},
	volume = {44},
	year = {1992}}

@article{PhysRevLett.131.046501,
  title = {Highly Resolved Spectral Functions of Two-Dimensional Systems with Neural Quantum States},
  author = {Mendes-Santos, Tiago and Schmitt, Markus and Heyl, Markus},
  journal = {Phys. Rev. Lett.},
  volume = {131},
  issue = {4},
  pages = {046501},
  numpages = {7},
  year = {2023},
  month = {Jul},
  publisher = {American Physical Society},
  doi = {10.1103/PhysRevLett.131.046501},
  url = {https://link.aps.org/doi/10.1103/PhysRevLett.131.046501}
}

@article{PhysRevB.105.205130,
  title = {Momentum-resolved time evolution with matrix product states},
  author = {Van Damme, Maarten and Vanderstraeten, Laurens},
  journal = {Phys. Rev. B},
  volume = {105},
  issue = {20},
  pages = {205130},
  numpages = {10},
  year = {2022},
  month = {May},
  publisher = {American Physical Society},
  doi = {10.1103/PhysRevB.105.205130},
  url = {https://link.aps.org/doi/10.1103/PhysRevB.105.205130}
}

@article{PhysRevLett.93.227205,
  title = {Density Matrix Renormalization Group and Periodic Boundary Conditions: A Quantum Information Perspective},
  author = {Verstraete, F. and Porras, D. and Cirac, J. I.},
  journal = {Phys. Rev. Lett.},
  volume = {93},
  issue = {22},
  pages = {227205},
  numpages = {4},
  year = {2004},
  month = {Nov},
  publisher = {American Physical Society},
  doi = {10.1103/PhysRevLett.93.227205},
  url = {https://link.aps.org/doi/10.1103/PhysRevLett.93.227205}
}

@article{RevModPhys.77.259,
  title = {The density-matrix renormalization group},
  author = {Schollw\"ock, U.},
  journal = {Rev. Mod. Phys.},
  volume = {77},
  issue = {1},
  pages = {259--315},
  numpages = {0},
  year = {2005},
  month = {Apr},
  publisher = {American Physical Society},
  doi = {10.1103/RevModPhys.77.259},
  url = {https://link.aps.org/doi/10.1103/RevModPhys.77.259}
}

@article{doi.org/10.1007/BF00655090,
	author = {Vidberg, H. J. and Serene, J. W.},
	date = {1977/11/01},
	doi = {10.1007/BF00655090},
	id = {Vidberg1977},
	isbn = {1573-7357},
	journal = {Journal of Low Temperature Physics},
	number = {3},
	pages = {179--192},
	title = {Solving the Eliashberg equations by means ofN-point Pad{\'e}approximants},
	url = {https://doi.org/10.1007/BF00655090},
	volume = {29},
	year = {1977}}

@book{10.7551/mitpress/3206.001.0001,
    author = {Rasmussen, Carl Edward and Williams, Christopher K. I.},
    title = {Gaussian Processes for Machine Learning},
    publisher = {The MIT Press},
    year = {2005},
    month = {11},
    isbn = {9780262256834},
    doi = {10.7551/mitpress/3206.001.0001},
    url = {https://doi.org/10.7551/mitpress/3206.001.0001},
    eprint = {https://direct.mit.edu/book-pdf/2514321/book\_9780262256834.pdf},
}

@article{scikit-learn,
  title={Scikit-learn: Machine Learning in {P}ython},
  author={Pedregosa, F. and Varoquaux, G. and Gramfort, A. and Michel, V.
          and Thirion, B. and Grisel, O. and Blondel, M. and Prettenhofer, P.
          and Weiss, R. and Dubourg, V. and Vanderplas, J. and Passos, A. and
          Cournapeau, D. and Brucher, M. and Perrot, M. and Duchesnay, E.},
  journal={Journal of Machine Learning Research},
  volume={12},
  pages={2825--2830},
  year={2011}
}

@misc{maxwellingkrr,
  author = {Welling Max},
  title = {Kernel ridge regression, Max Welling’s classnotes in machine learning.},
  year = {2013}
}

@article{PhysRevLett.73.732,
  title = {Spectral Properties of the One-Dimensional Hubbard Model},
  author = {Preuss, R. and Muramatsu, A. and von der Linden, W. and Dieterich, P. and Assaad, F. F. and Hanke, W.},
  journal = {Phys. Rev. Lett.},
  volume = {73},
  issue = {5},
  pages = {732--735},
  numpages = {0},
  year = {1994},
  month = {Aug},
  publisher = {American Physical Society},
  doi = {10.1103/PhysRevLett.73.732},
  url = {https://link.aps.org/doi/10.1103/PhysRevLett.73.732}
}

@article{PhysRevB.109.045102,
  title = {Spectral properties of a one-dimensional extended Hubbard model from bosonization and time-dependent variational principle: Applications to one-dimensional cuprates},
  author = {Wang, Hao-Xin and Wu, Yi-Ming and Jiang, Yi-Fan and Yao, Hong},
  journal = {Phys. Rev. B},
  volume = {109},
  issue = {4},
  pages = {045102},
  numpages = {8},
  year = {2024},
  month = {Jan},
  publisher = {American Physical Society},
  doi = {10.1103/PhysRevB.109.045102},
  url = {https://link.aps.org/doi/10.1103/PhysRevB.109.045102}
}

@article{PhysRevB.71.104520,
  title = {Superconductivity in the quasi-two-dimensional Hubbard model},
  author = {Yan, Xin-Zhong},
  journal = {Phys. Rev. B},
  volume = {71},
  issue = {10},
  pages = {104520},
  numpages = {16},
  year = {2005},
  month = {Mar},
  publisher = {American Physical Society},
  doi = {10.1103/PhysRevB.71.104520},
  url = {https://link.aps.org/doi/10.1103/PhysRevB.71.104520}
}

@article{PhysRevX.8.021048,
  title = {Pseudogap and Fermi-Surface Topology in the Two-Dimensional Hubbard Model},
  author = {Wu, Wei and Scheurer, Mathias S. and Chatterjee, Shubhayu and Sachdev, Subir and Georges, Antoine and Ferrero, Michel},
  journal = {Phys. Rev. X},
  volume = {8},
  issue = {2},
  pages = {021048},
  numpages = {17},
  year = {2018},
  month = {May},
  publisher = {American Physical Society},
  doi = {10.1103/PhysRevX.8.021048},
  url = {https://link.aps.org/doi/10.1103/PhysRevX.8.021048}
}

@article{RevModPhys.68.13,
  title = {Dynamical mean-field theory of strongly correlated fermion systems and the limit of infinite dimensions},
  author = {Georges, Antoine and Kotliar, Gabriel and Krauth, Werner and Rozenberg, Marcelo J.},
  journal = {Rev. Mod. Phys.},
  volume = {68},
  issue = {1},
  pages = {13--125},
  numpages = {0},
  year = {1996},
  month = {Jan},
  publisher = {American Physical Society},
  doi = {10.1103/RevModPhys.68.13},
  url = {https://link.aps.org/doi/10.1103/RevModPhys.68.13}
}

@article{10.1063/1.1712502,
    author = {Kotliar, Gabriel and Vollhardt, Dieter},
    title = {Strongly Correlated Materials: Insights From Dynamical Mean-Field Theory},
    journal = {Physics Today},
    volume = {57},
    number = {3},
    pages = {53-59},
    year = {2004},
    month = {03},
    issn = {0031-9228},
    doi = {10.1063/1.1712502},
    url = {https://doi.org/10.1063/1.1712502},
    eprint = {https://pubs.aip.org/physicstoday/article-pdf/57/3/53/11004005/53_1_online.pdf},
}

@article{RevModPhys.78.865,
  title = {Electronic structure calculations with dynamical mean-field theory},
  author = {Kotliar, G. and Savrasov, S. Y. and Haule, K. and Oudovenko, V. S. and Parcollet, O. and Marianetti, C. A.},
  journal = {Rev. Mod. Phys.},
  volume = {78},
  issue = {3},
  pages = {865--951},
  numpages = {0},
  year = {2006},
  month = {Aug},
  publisher = {American Physical Society},
  doi = {10.1103/RevModPhys.78.865},
  url = {https://link.aps.org/doi/10.1103/RevModPhys.78.865}
}

@article{PhysRevB.102.081110,
  title = {The Mott transition as a topological phase transition},
  author = {Sen, Sudeshna and Wong, Patrick J. and Mitchell, Andrew K.},
  journal = {Phys. Rev. B},
  volume = {102},
  issue = {8},
  pages = {081110},
  numpages = {6},
  year = {2020},
  month = {Aug},
  publisher = {American Physical Society},
  doi = {10.1103/PhysRevB.102.081110},
  url = {https://link.aps.org/doi/10.1103/PhysRevB.102.081110}
}

@article{PhysRevB.90.155136,
  title = {Machine learning for many-body physics: The case of the Anderson impurity model},
  author = {Arsenault, Louis-Fran\ifmmode \mbox{\c{c}}\else \c{c}\fi{}ois and Lopez-Bezanilla, Alejandro and von Lilienfeld, O. Anatole and Millis, Andrew J.},
  journal = {Phys. Rev. B},
  volume = {90},
  issue = {15},
  pages = {155136},
  numpages = {16},
  year = {2014},
  month = {Oct},
  publisher = {American Physical Society},
  doi = {10.1103/PhysRevB.90.155136},
  url = {https://link.aps.org/doi/10.1103/PhysRevB.90.155136}
}

@article{PhysRevB.91.155107,
  title = {Spectral functions of strongly correlated extended systems via an exact quantum embedding},
  author = {Booth, George H. and Chan, Garnet Kin-Lic},
  journal = {Phys. Rev. B},
  volume = {91},
  issue = {15},
  pages = {155107},
  numpages = {6},
  year = {2015},
  month = {Apr},
  publisher = {American Physical Society},
  doi = {10.1103/PhysRevB.91.155107},
  url = {https://link.aps.org/doi/10.1103/PhysRevB.91.155107}
}

@article{PhysRevB.88.041107,
  title = {Limitations of the hybrid functional approach to electronic structure of transition metal oxides},
  author = {Coulter, John E. and Manousakis, Efstratios and Gali, Adam},
  journal = {Phys. Rev. B},
  volume = {88},
  issue = {4},
  pages = {041107},
  numpages = {5},
  year = {2013},
  month = {Jul},
  publisher = {American Physical Society},
  doi = {10.1103/PhysRevB.88.041107},
  url = {https://link.aps.org/doi/10.1103/PhysRevB.88.041107}
}

@article{RevModPhys.79.291,
  title = {Coupled-cluster theory in quantum chemistry},
  author = {Bartlett, Rodney J. and Musia\l{}, Monika},
  journal = {Rev. Mod. Phys.},
  volume = {79},
  issue = {1},
  pages = {291--352},
  numpages = {0},
  year = {2007},
  month = {Feb},
  publisher = {American Physical Society},
  doi = {10.1103/RevModPhys.79.291},
  url = {https://link.aps.org/doi/10.1103/RevModPhys.79.291}
}

@article{doi:10.1021/cr200107z,
	author = {Cohen, Aron J. and Mori-S{\'a}nchez, Paula and Yang, Weitao},
	date = {2012/01/11},
	doi = {10.1021/cr200107z},
	isbn = {0009-2665},
	journal = {Chemical Reviews},
	journal1 = {Chemical Reviews},
	journal2 = {Chem. Rev.},
	month = {01},
	number = {1},
	pages = {289--320},
	publisher = {American Chemical Society},
	title = {Challenges for Density Functional Theory},
	type = {doi: 10.1021/cr200107z},
	url = {https://doi.org/10.1021/cr200107z},
	volume = {112},
	year = {2012},
	year1 = {2012}}

@article{Zhu_2025,
	author = {Zhu, Yuanran and Yin, Jia and Reeves, Cian C and Yang, Chao and Vl{\v c}ek, Vojt{\v e}ch},
	doi = {10.1088/2632-2153/ada99d},
	journal = {Machine Learning: Science and Technology},
	month = {feb},
	number = {1},
	pages = {015027},
	publisher = {IOP Publishing},
	title = {Predicting nonequilibrium Green's function dynamics and photoemission spectra via nonlinear integral operator learning},
	url = {https://doi.org/10.1088/2632-2153/ada99d},
	volume = {6},
	year = {2025}}

@article{PhysRevB.103.245118,
  title = {Predicting impurity spectral functions using machine learning},
  author = {Sturm, Erica J. and Carbone, Matthew R. and Lu, Deyu and Weichselbaum, Andreas and Konik, Robert M.},
  journal = {Phys. Rev. B},
  volume = {103},
  issue = {24},
  pages = {245118},
  numpages = {14},
  year = {2021},
  month = {Jun},
  publisher = {American Physical Society},
  doi = {10.1103/PhysRevB.103.245118},
  url = {https://link.aps.org/doi/10.1103/PhysRevB.103.245118}
}

@misc{supp,
  note = "See Supplemental Material at
     [URL will be inserted by publisher] for an overview of the DOS obtained from HF and first-order $G_0W_0$ calculations and many-body predictions, additional data on the DOS error induced by analytic continuation, a comparison of different kernel functions, and an overview of the ARD for long-range hopping terms with $t'''=0.1$."
}

@article{doi.org/10.1038/s43588-025-00810-z,
	author = {Venturella, Christian and Li, Jiachen and Hillenbrand, Christopher and Leyva Peralta, Ximena and Liu, Jessica and Zhu, Tianyu},
	date = {2025/06/01},
	doi = {10.1038/s43588-025-00810-z},
	id = {Venturella2025},
	isbn = {2662-8457},
	journal = {Nature Computational Science},
	number = {6},
	pages = {502--513},
	title = {Unified deep learning framework for many-body quantum chemistry via Green's functions},
	url = {https://doi.org/10.1038/s43588-025-00810-z},
	volume = {5},
	year = {2025}}

@article{10.5555/944790.944815, author = {Genton, Marc G.}, title = {Classes of kernels for machine learning: a statistics perspective}, year = {2002}, issue_date = {3/1/2002}, publisher = {JMLR.org}, volume = {2}, issn = {1532-4435}, journal = {J. Mach. Learn. Res.}, month = mar, pages = {299–312}, numpages = {14} }

@article{https://doi.org/10.1038/s41467-024-53748-7,
	author = {Hou, Bowen and Wu, Jinyuan and Qiu, Diana Y.},
	date = {2024/11/02},
	doi = {10.1038/s41467-024-53748-7},
	id = {Hou2024},
	isbn = {2041-1723},
	journal = {Nature Communications},
	number = {1},
	pages = {9481},
	title = {Unsupervised representation learning of Kohn--Sham states and consequences for downstream predictions of many-body effects},
	url = {https://doi.org/10.1038/s41467-024-53748-7},
	volume = {15},
	year = {2024}}

@article{PhysRevB.107.075151,
  title = {Robust analytic continuation of Green's functions via projection, pole estimation, and semidefinite relaxation},
  author = {Huang, Zhen and Gull, Emanuel and Lin, Lin},
  journal = {Phys. Rev. B},
  volume = {107},
  issue = {7},
  pages = {075151},
  numpages = {13},
  year = {2023},
  month = {Feb},
  publisher = {American Physical Society},
  doi = {10.1103/PhysRevB.107.075151},
  url = {https://link.aps.org/doi/10.1103/PhysRevB.107.075151}
}

@article{PhysRevB.106.235106,
  title = {Single- and two-particle finite size effects in interacting lattice systems},
  author = {Iskakov, Sergei and Terletska, Hanna and Gull, Emanuel},
  journal = {Phys. Rev. B},
  volume = {106},
  issue = {23},
  pages = {235106},
  numpages = {12},
  year = {2022},
  month = {Dec},
  publisher = {American Physical Society},
  doi = {10.1103/PhysRevB.106.235106},
  url = {https://link.aps.org/doi/10.1103/PhysRevB.106.235106}
}

@article{RevModPhys.77.1027,
  title = {Quantum cluster theories},
  author = {Maier, Thomas and Jarrell, Mark and Pruschke, Thomas and Hettler, Matthias H.},
  journal = {Rev. Mod. Phys.},
  volume = {77},
  issue = {3},
  pages = {1027--1080},
  numpages = {0},
  year = {2005},
  month = {Oct},
  publisher = {American Physical Society},
  doi = {10.1103/RevModPhys.77.1027},
  url = {https://link.aps.org/doi/10.1103/RevModPhys.77.1027}
}

@article{doi.org/10.1140/epjs/s11734-023-00976-5,
	author = {Enenkel, Nicklas and Garst, Markus and Schmitteckert, Peter},
	date = {2023/12/01},
	doi = {10.1140/epjs/s11734-023-00976-5},
	id = {Enenkel2023},
	isbn = {1951-6401},
	journal = {The European Physical Journal Special Topics},
	number = {20},
	pages = {3495--3504},
	title = {Applicability and limitations of cluster perturbation theory for Hubbard models},
	url = {https://doi.org/10.1140/epjs/s11734-023-00976-5},
	volume = {232},
	year = {2023}}
\bibliographystyle{unsrt}

\newpage
\onecolumngrid
\appendix
\section{}
\label{sec:appendix}
\counterwithin{figure}{section}

\subsection{Hartree-Fock Convergence}
In the following, we present our procedure to find the Hartree-Fock (HF) ground state of the $L=10$ Hubbard model. In order to reach convergence, we use an iterative approach with an initial density matrix guess for all calculations over the range of $U$-values with $N$ electrons: 
\begin{equation}
    \rho_{\text{init}} = \frac{1}{2}\Big(\rho_{\text{alt}} + \rho_{\text{prev}}\Big)
\end{equation}
where $\rho_{\text{prev}}$ is the converged density matrix of the next-smallest $U$-value, and $\rho_{\text{prev}} = \rho_{\text{alt}}$ for the smallest $U$-value. $\rho_{\text{alt}} = (\rho^{\uparrow}_{\text{alt}},\rho^{\downarrow}_{\text{alt}})$ consists of $\rho^{\uparrow}_{\text{alt}} = \text{diag}(0.7, 0.3, ...)$ and $\rho^{\downarrow}_{\text{alt}} = \text{diag}(0.3, 0.7, ...)$ such that in total $N$ elements of $\rho_{\text{alt}}$ are nonzero. If this calculation does not converge, we first use only the previous density matrix $\rho_{\text{init}} = \rho_{\text{prev}}$ and then add a random initial guess:
\begin{equation}
    \rho_{\text{init}} = a \cdot \rho_{\text{random}} + b \cdot \rho_{\text{prev}}
\end{equation}
with 
\begin{equation}
    \rho_{\text{random}} = (\text{diag}(x_1, ..., x_L),\text{diag}(x'_1, ..., x'_L))
\end{equation} 
where $\{x_i\}_{i = 1,...,L} \,, \{x'_i\}_{i = 1,...,L}$ are independent random numbers. We define a maximum number of iterations $M$ and determine the mixture of the previous density matrix $\rho_{\text{prev}}$ and the random density matrix $\rho_{\text{random}}$ in the $j$-th round through
\begin{equation}
    r = \frac{M-2- (j-2))}{M - 2}\,,\quad a = 1-r\,,\quad b = r \,.
\end{equation}
For the $M/2$-th round, we add random off-diagonal elements: 
\begin{equation}
    \rho_{\text{random}} = (X,X')
\end{equation}where $\{X_{ij}\}_{ij = 1,...,L} \,, \{X'_{i,j}\}_{i,j = 1,...,L}$ are random matrix elements. This procedure often converges within the first iteration, and we typically need less than 10 iterations for most parameters.

\subsection{Machine-Learning Method}
\begin{figure}[htbp!]
 \centering
 \begin{adjustbox}{center}
   \includegraphics[width=1\columnwidth]{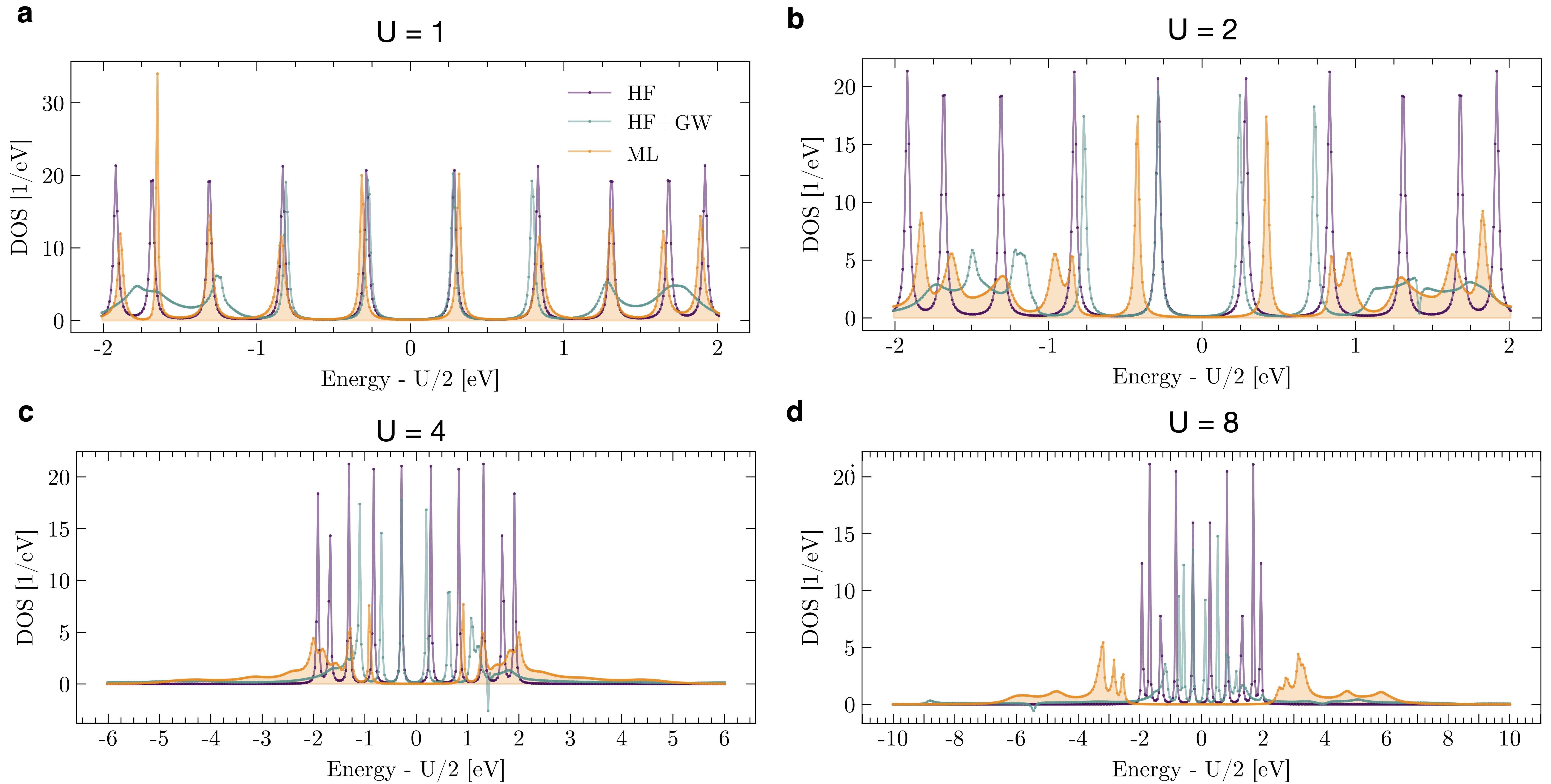}
 \end{adjustbox}
 \caption{Difference in DOS of a nearest-neighbor Hubbard model between the mean-field Green's function (HF, purple) and HF-GF with self-energy from first order \textit{GW}-calculations (HF + GW, turquoise) in comparison to the predicted Green's function (ML, orange) for different Coulomb repulsions: (a) $U=1$. (b) $U=2$. (c) $U=4$. (d) $U=8$.}
 \label{fig:fig_SI1B}
\end{figure}

One may argue that spectral functions can easily be obtained from computationally less expensive methods than FCI, hence making the ML approach redundant. While this is certainly the case for weakly interacting systems such as molecules, we show in fig. \ref{fig:fig_SI1B}(a) that the DOS is different even at $U = 1$ if obtained only from mean-field calculations (purple) compared to our ML method which incorporates the many-body interactions (orange). In the former case, we see that the quasiparticle peaks are shifted towards the Fermi level. We also compare a calculation which uses the first-order $G_0 W_0$ self-energy combined with the mean-field Green's function $G_0$ via Dyson's equation (turquoise). This data shows that we are not providing the many-body self-energy to our ML model by using the $G_0 W_0$ self-energy $\Sigma_{\text{GW}}$ in our feature vector. In fig. \ref{fig:fig_SI1B}(b), (c), and (d), we show the same data for $U=2$, $U=4$, and $U=8$, respectively. We can clearly see that the DOS is not accurate at all if we would only rely on a mean-field framework.  

The ML-GF is obtained by predicting the self-energy on the imaginary frequency axis, combining it with the HF-GF to obtain the predicted Green's function on the imaginary axis, and then using A.C. to obtain the ML-GF from which we can calculate the DOS as a function of energy. We found that reversing the steps, i.e., first performing A.C. and then using Dyson's equation, consistently led to noisier results. We also performed calculations with different linear combinations of Laplacian and Gaussian kernels, as well as combinations of these with Mátern functions, but we did not find consistent trends to further increase accuracy. 

\subsection{DOS Error}

\begin{figure}[htbp!]
 \centering
 \begin{adjustbox}{center}
   \includegraphics[width=1\columnwidth]{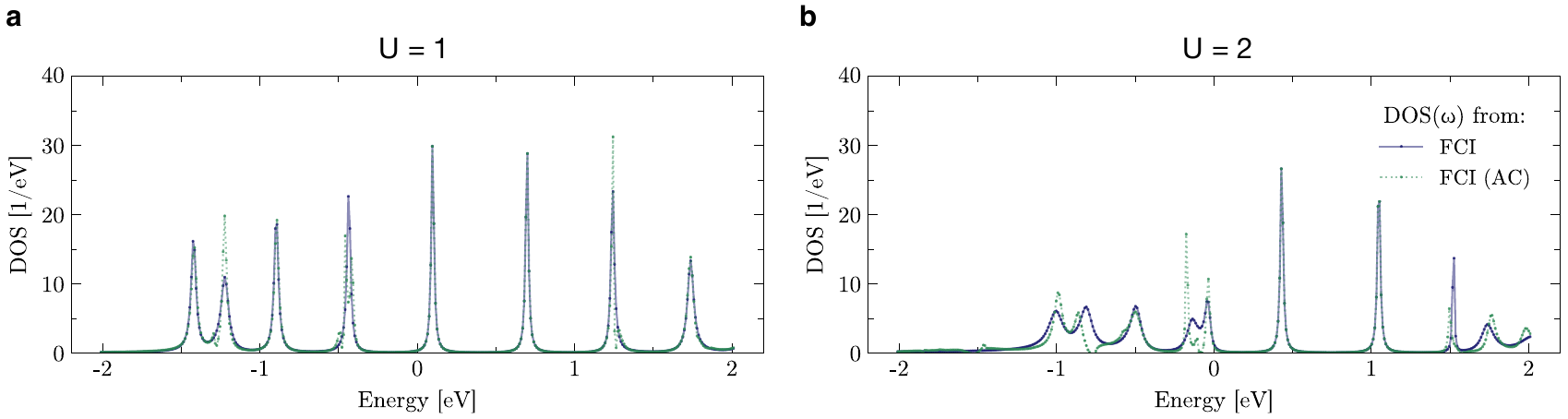}
 \end{adjustbox}
 \caption{DOS of a nearest-neighbor Hubbard model. Comparison of exact solution (FCI, blue) to exact solution with additional A.C. steps (FCI (AC), green) for (a) $U = 1$ and (b) $U = 2$.}
 \label{fig:fig_SI9}
\end{figure}

In the following, we provide additional evidence supporting our attribution of the dominant error in the DOS to the A.C. step in our workflow.
In fig. \ref{fig:fig_SI9}, we present the DOS of a nearest-neighbor Hubbard model for (a) $U=1$ and (b) $U=2$. We compare the DOS obtained directly from the exact FCI Green’s function (labeled FCI): as
\begin{equation}
    \underbrace{G(i\omega) \longrightarrow  G(\omega + i\eta)}_{\text{A.C.}}
\end{equation}
with the DOS obtained after applying additional analytic continuation steps to otherwise exact imaginary-frequency data, labeled FCI (AC), where both the self-energy and the HF Green's function are first analytically continued using the Padé method. Specifically, starting from the FCI Green’s function on the imaginary axis $G(i\omega)$ and the HF-GF $G_0(i\omega)$, we obtain the imaginary-frequency self-energy:
\begin{equation}
    \Sigma(i\omega) = G_0(i\omega)^{-1} - G(i\omega)^{-1}
\end{equation}
Then, we perform A.C.:
\begin{equation}
    \underbrace{\Sigma(i\omega) \longrightarrow  \Sigma(\omega + i\eta)}_{\text{A.C.}} \quad\text{and}\quad  \underbrace{G_0(i\omega) \longrightarrow G_0(\omega + i\eta)}_{\text{A.C.}} 
\end{equation}
and reconstruct the real-frequency Green’s function using Dyson’s equation:
\begin{equation}
    G(\omega+i\eta) = \left( G_0(\omega+i\eta)^{-1} - \Sigma(\omega+i\eta) \right)^{-1} \,.
\end{equation}
Despite the fact that the underlying imaginary-frequency data are the same, we observe that the additional A.C. procedure introduces noticeable deviations from the reference DOS (FCI). This demonstrates that a substantial portion of the DOS error arises even in the absence of any ML-related errors.

\begin{figure}[htbp!]
 \centering
 \begin{adjustbox}{center}
   \includegraphics[width=1\columnwidth]{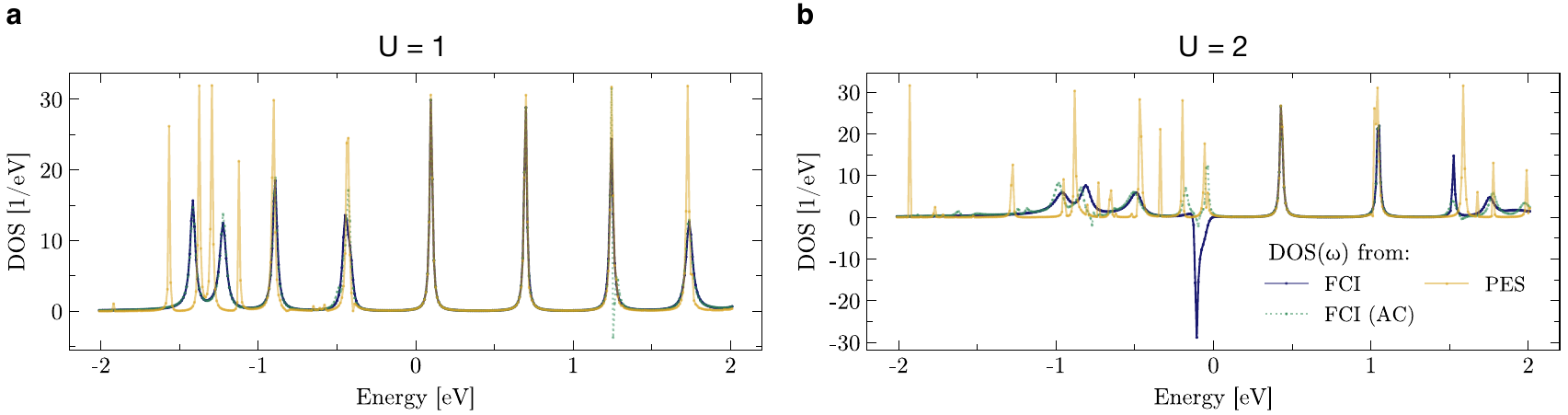}
 \end{adjustbox}
 \caption{DOS of a nearest-neighbor Hubbard model. Comparison of exact solution (FCI, blue) to exact solution with additional A.C. steps (FCI AC, green) and PES A.C. method (PES, yellow) for (a) $U = 1$ and (b) $U = 2$.}
 \label{fig:fig_SI10}
\end{figure}

Figure \ref{fig:fig_SI10} extends this analysis by comparing two different analytic continuation schemes for the same Hamiltonian parameters presented in fig. \ref{fig:fig_SI9}. In addition to the Padé A.C. (FCI AC), we include results obtained using the pole-expansion scheme (PES) analytic continuation method. While the detailed spectral line shapes depend on the respective A.C. algorithm, both methods yield DOS variations of comparable magnitude. This indicates that the error in the reconstructed real-frequency spectra is dominated by the intrinsic instability of analytic continuation, rather than by the specific continuation method employed.

These results provide evidence that the dominant source of error in the DOS originates from the A.C. step itself. 

\subsection{Accuracy of Machine-Learning Predictions}
\begin{figure}[htbp!]
 \centering
 \begin{adjustbox}{center}
   \includegraphics[width=1\columnwidth]{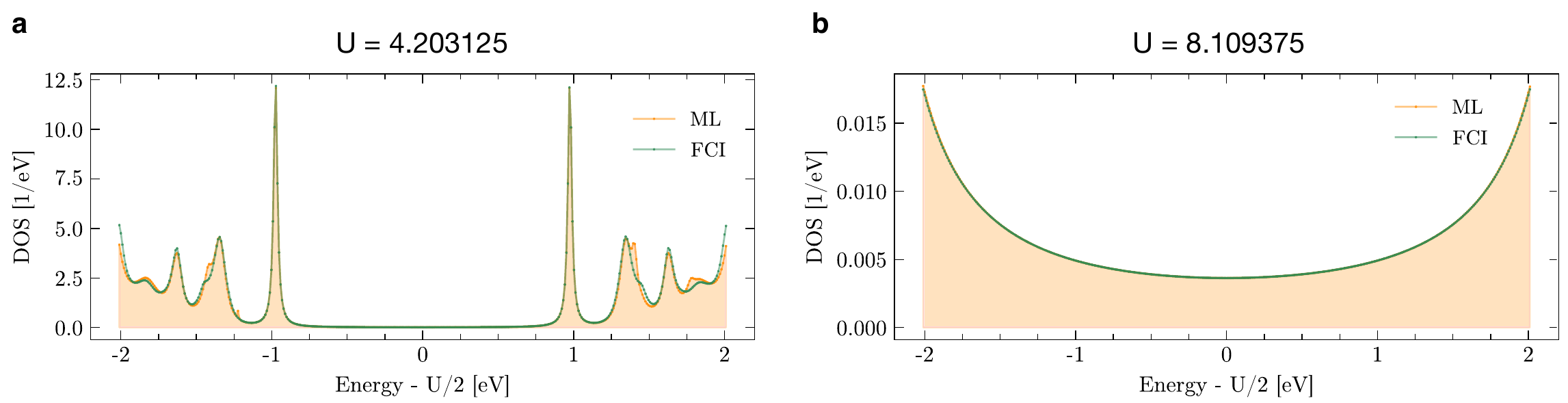}
 \end{adjustbox}
 \caption{DOS of a nearest-neighbor Hubbard model. Comparison of prediction (ML, orange) and exact solution (FCI, green) for training data with (a) $U = 4.203125$ and (b) $U = 8.109375$.}
 \label{fig:fig_SI2B}
\end{figure}

In fig. \ref{fig:fig_SI2B}, we show that the ML predictions of the DOS for (a) $U \approx 4$ and (b) $U \approx 8$ align well with the exact solution close to the Fermi edge, while the main deviations occur far away from it.

\begin{figure*}[htbp!]
 \centering
 \begin{adjustbox}{center}
   \includegraphics[width=1\columnwidth]{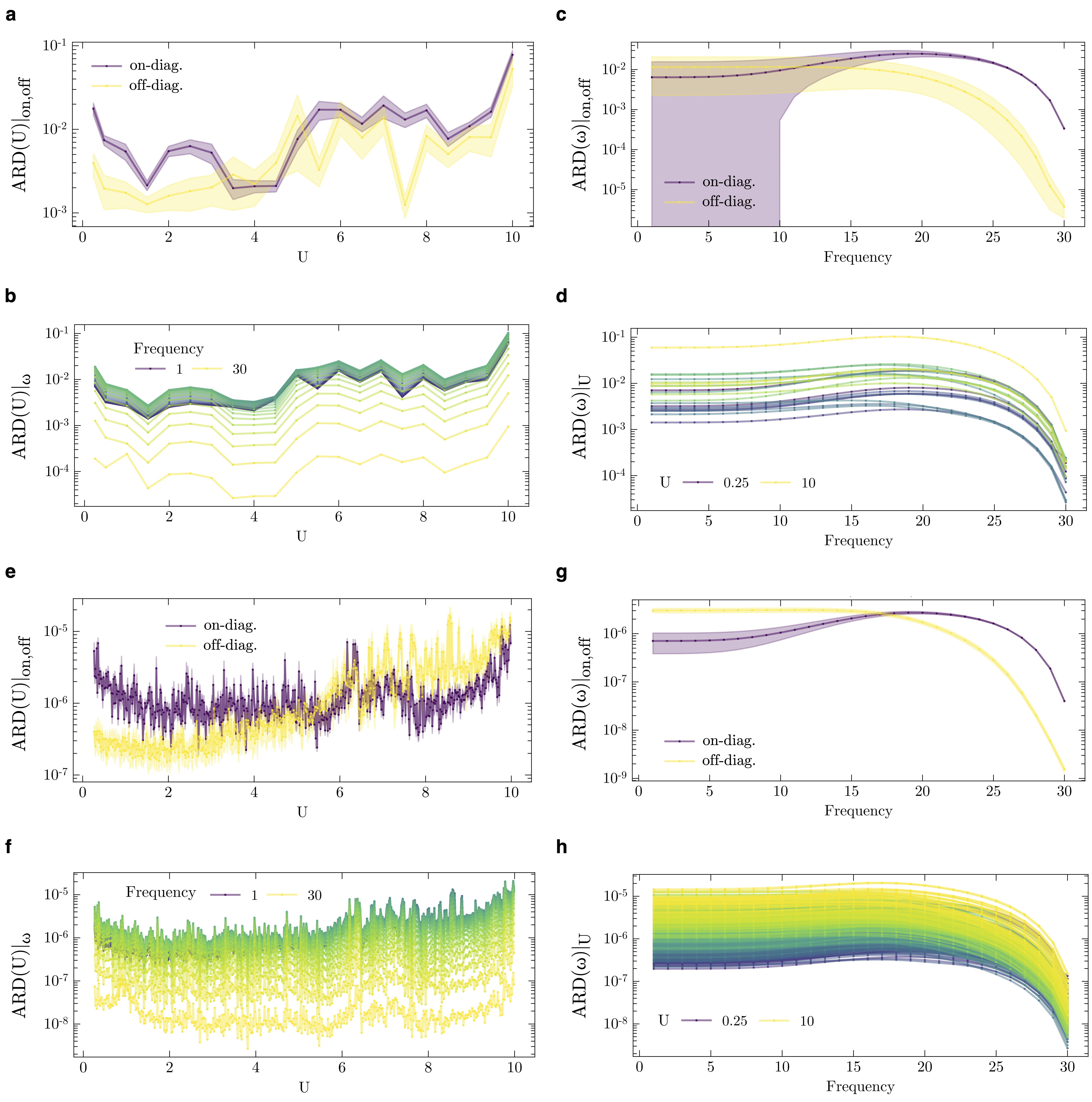}
 \end{adjustbox}
    \caption{Overview of ARD between exact and predicted self-energy for a Hubbard model with long-range interactions $t' = 0.25$, $t''=0.1$, $t''' = 0.1$, and with kernel $K_6$. ARD for (a)-(d) test data and (e)-(h) training data.}
 \label{fig:fig6}
\end{figure*}

Our ML framework reproduces training data with high accuracy, both for nearest-neighbor Hubbard models and for Hubbard models with long-range hopping terms.
In fig. \ref{fig:fig6}, we present the same overview of the ARD for the case with $t =1$, $t'=0.25$, $t''=0.1$, and $t''' = 0.1$, including test data, see fig. \ref{fig:fig6}(a)-(d), and training data in fig. \ref{fig:fig6}(e)-(h). Note that we used a kernel with six Matérn functions to achieve high accuracy. For the test data, we find a similar ARD compared to the model with $t'''=0$, see fig. \ref{fig:fig6}(a). However, the training data reveals that this system can be learned with slightly higher precision. The highest ARD in fig. \ref{fig:fig6}(e) is around $2 \cdot 10^{-5}$, while we find approximately $7 \cdot 10^{-5}$ for $t''' = 0$. More remarkably, we find a large difference between the Hubbard model with nearest-neighbor interactions $t=1$ and the case with additional hopping terms $t'$, $t''$ and $t'''$, see fig. \ref{fig:fig6}(e). In the former case, we find ARD values between $10^{-6}$ and $10^{-4}$ for on-diagonal elements and $10^{-6}$ and $10^{-1}$ for off-diagonal elements. On the other hand, the latter model with long-range terms leads to more accurate predictions with ARD values between $10^{-7}$ and $10^{-5}$ for both on- and off-diagonal elements. 

\begin{figure}[htbp!]
 \centering
 \begin{adjustbox}{center}
   \includegraphics[width=1\columnwidth]{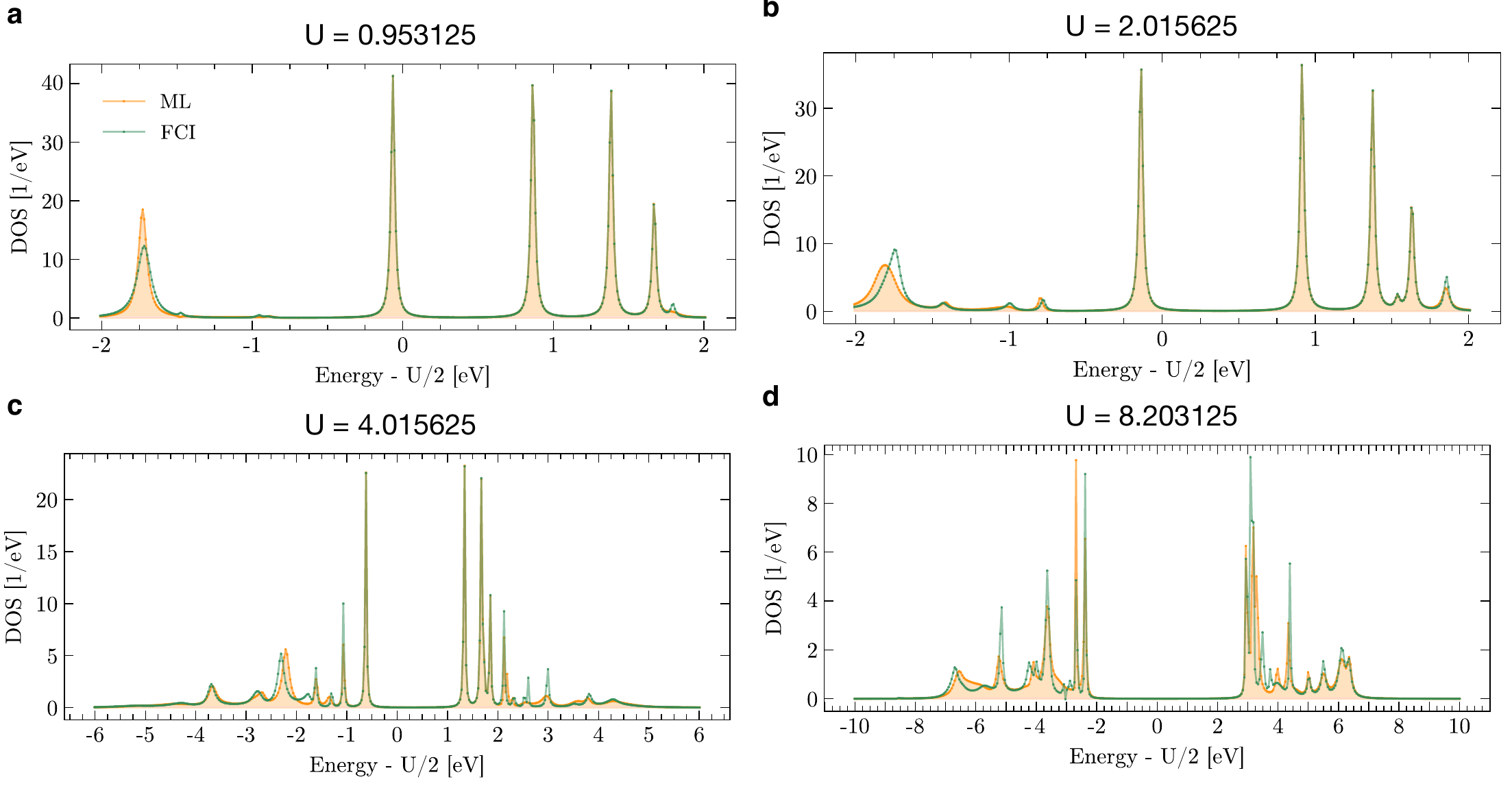}
 \end{adjustbox}
 \caption{DOS of a Hubbard model with long-range hopping terms $t = 1$, $t' = 0.25$, and $t'' = 0.1$. Comparison of ML prediction (orange) and FCI (green) for training data with different Coulomb repulsions (a) $U = 0.953125$, (b) $U = 2.015625$, (c) $U=4.015625$, and (d) $U = 8.203125$. }
 \label{fig:fig_SI4C}
\end{figure}

\begin{figure}[htbp!]
 \centering
 \begin{adjustbox}{center}
   \includegraphics[width=1\columnwidth]{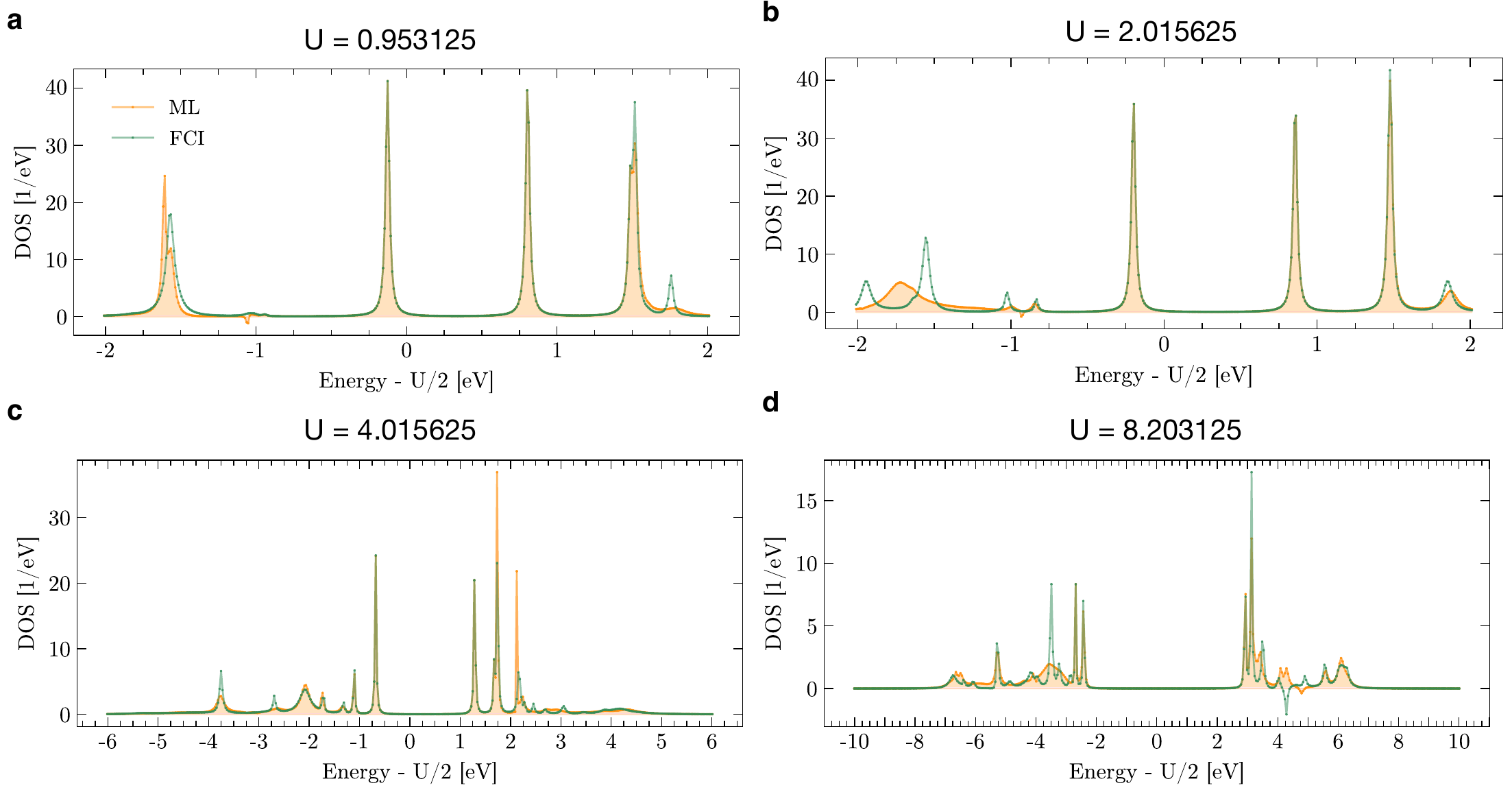}
 \end{adjustbox}
 \caption{Same as figure \ref{fig:fig_SI4C}, but for training data of a Hubbard model with long-range hopping $t = 1$, $t' = 0.25$, $t'' = 0.1$, and $t''' = 0.1$.}
 \label{fig:fig_SI4E}
\end{figure}

In the following, we show that these self-energy predictions also lead to an accurate DOS. In fig. \ref{fig:fig_SI4C}, we present the ML prediction for training data (orange area) and the exact FCI solution (green line) for a Hubbard model with second ($t' = 0.25$) and third ($t'' = 0.1$) neighbor hopping terms and Coulomb repulsions close to $U \approx 1$, $U \approx 2$, $U \approx 4$, and $U \approx 8$, see fig. \ref{fig:fig_SI4C}(a)-(d), respectively. In fig. \ref{fig:fig_SI4E}, we present the DOS prediction for a Hubbard model which also has an additional hopping $t''' = 0.1$ for the same values of $U \approx 1$, $U \approx 2$, $U \approx 4$, and $U \approx 8$. In this case, we use a kernel with six Matérn functions ($K_6$) instead of three. Even for large $U = 8.203125$, the model captures the many-body peaks and their approximate density around $\pm 7$ eV from $E - E_F$. Note that the Fermi level is different from the nearest-neighbor case $E_F = U/2$.

\subsection{Comparison of Kernel Functions}
\begin{figure*}[htbp!]
 \centering
 \begin{adjustbox}{center}
   \includegraphics[width=1\columnwidth]{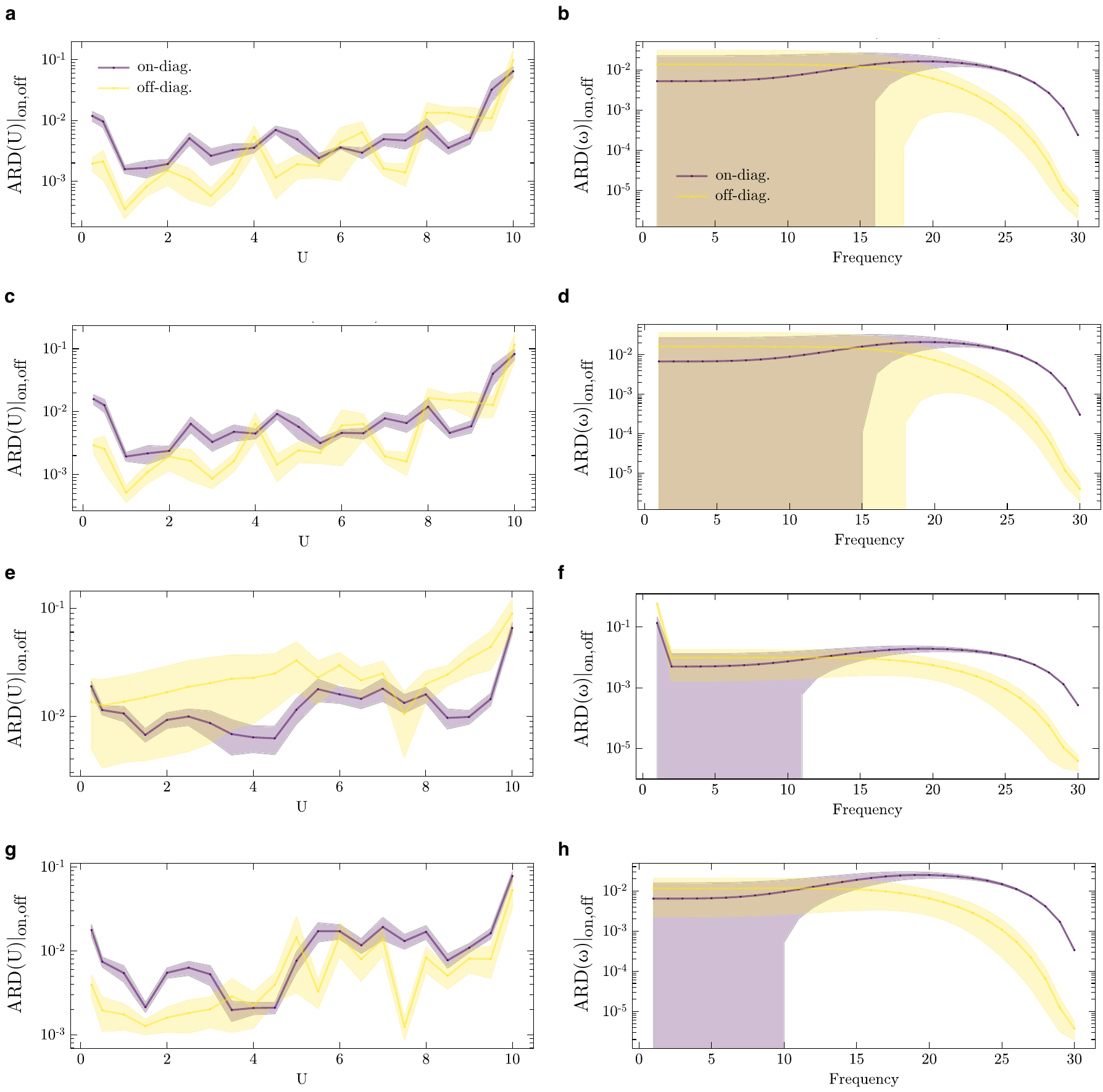}
 \end{adjustbox}
    \caption{ARD for test data of a Hubbard model with long-range hopping terms $t = 1$, $t' = 0.25$, $t'' = 0.1$ and (a)-(d) $t''' = 0$ or (e)-(h) $t''' = 0.1$. We compare a kernel with (a)-(b) three Matérn functions $K_3$ to (c)-(d) six Matérn function $K_6$ for $t''' = 0$. Left column: ARD for all on- (purple line) and off-diagonal (yellow line) matrix elements for different Coulomb interactions $U$. Right column: ARD as a function of frequency. Shaded areas indicate the standard deviation over all matrix elements of different frequencies (left) or $U$-values (right). (e)-(h) Same calculation for $t''' = 0.1$ with kernels (e)-(f) $K_3$ and (g)-(h) $K_6$.}
 \label{fig:figureSI5_6A}
\end{figure*}

In fig. \ref{fig:figureSI5_6A}, we present the ARD of our self-energy predictions for unseen test data for a Hubbard model with long-range hopping terms $t' = 0.25$, $t'' =0.1$ and (a)-(d) $t''' = 0$ as well as (e)-(h) $t''' = 0.1$. We compare the ARD for on- and off-diagonal elements as a function of $U$-value and frequency in fig. \ref{fig:figureSI5_6A}(a) and (b), respectively, with the same metric in fig. \ref{fig:figureSI5_6A}(c)-(d), where we use six instead of three Matérn functions in our kernel. We observe that there is no noticeable difference between these two calculations, indicating that three Matérn functions are sufficient to capture all important relationships between the features and the self-energy target for the Hubbard model with $t''' = 0$. On the other hand, we can see from fig. \ref{fig:figureSI5_6A}(e) and (f) compared to fig. \ref{fig:figureSI5_6A}(g) and (h), respectively, that the accuracy increases significantly in the Hubbard model with $t'''\neq0$ when using a kernel with six Matérn functions. This includes both the standard deviation, visualized as the shaded area, as well as the ARD value itself. We want to emphasize the first frequency index $\omega_1$ has a significantly higher ARD compared to all other frequency values when using only three Matérn function, see fig. \ref{fig:figureSI5_6A}(f), which can be improved by using six Matérn functions, as shown in fig. \ref{fig:figureSI5_6A}(h). 

\begin{figure*}[htbp!]
 \centering
 \begin{adjustbox}{center}
   \includegraphics[width=1\columnwidth]{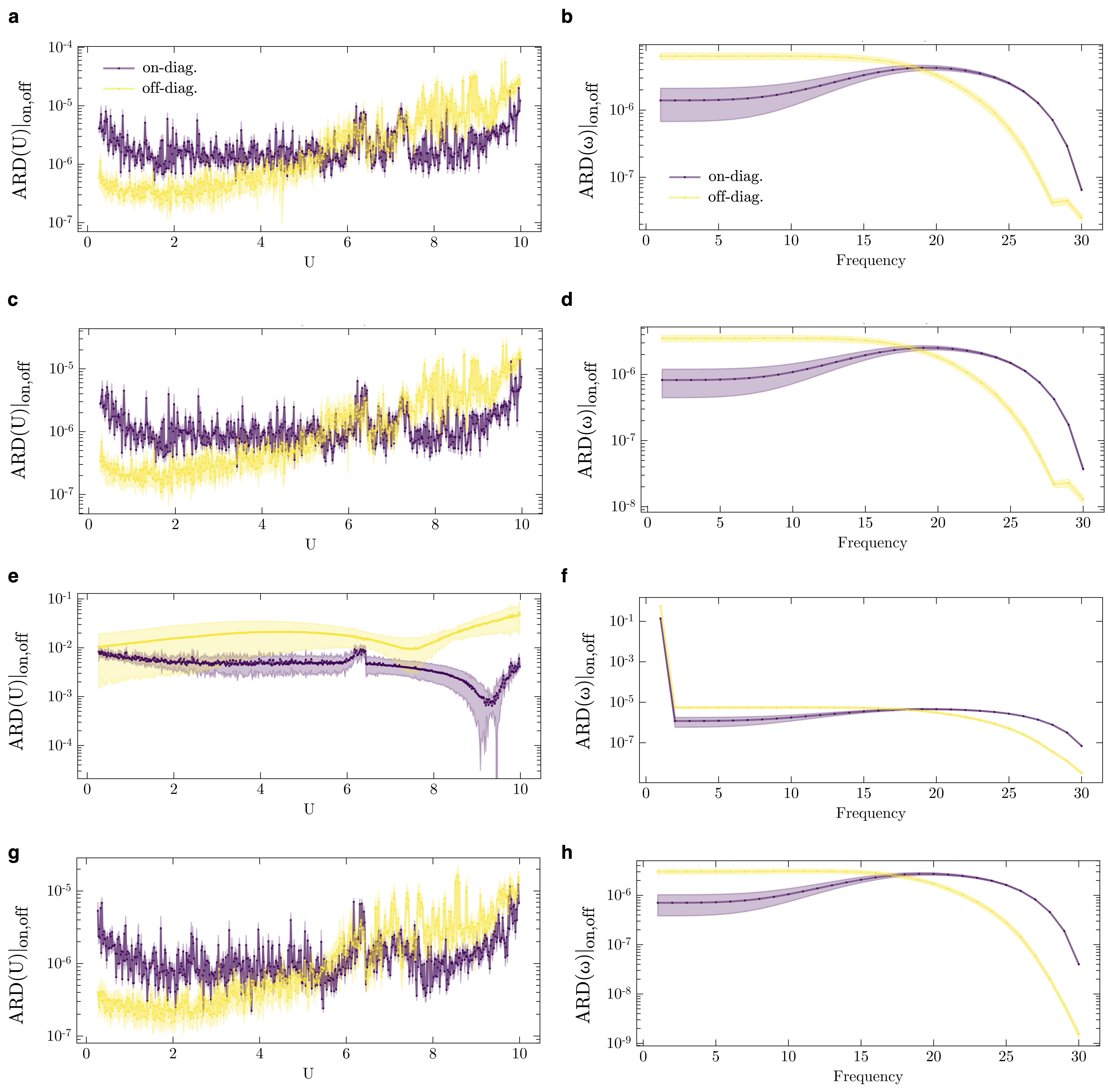}
 \end{adjustbox}
    \caption{Same as fig. \ref{fig:figureSI5_6A}, but for training data of a Hubbard model with $t = 1$, $t' = 0.25$, $t'' = 0.1$ and (a)-(d) $t''' = 0$ or (e)-(h) $t''' = 0.1$. Comparison of (a)-(b) and (e)-(f) three Matérn functions $K_3$ to (c)-(d) and (g)-(h) six Matérn functions $K_6$.}
 \label{fig:figureSI5_6B}
\end{figure*}

The same increase in prediction accuracy can also be observed for training data, see fig. \ref{fig:figureSI5_6B}(a)-(d) for $t''' = 0$ and (e)-(h) for $t''' = 0.1$. There is only a slight increase in accuracy between three and six Matérn functions for $t''' = 0$, see fig. \ref{fig:figureSI5_6B}(a) versus (c) and also (b) versus (d). For the Hubbard model with $t''' = 0.1$, we observe a difference on the first frequency index $\omega_1$, see fig. \ref{fig:figureSI5_6B}(f) compared to (h). In fig. \ref{fig:figureSI5_6B}(e)-(f) versus (g)-(h), respectively, we observe that using more Matérn functions solves this issue, leading to a lower ARD. 

\end{document}